\documentclass[epsf,aps,twocolumn,showpacs,preprintnumbers,amsmath,amssymb,footinbib]{revtex4}
\usepackage{epsf}
\usepackage{hyperref}
\newcommand{\beq}{\begin{equation}}

\newcommand{\boldl}{{\bf L}}
\newcommand{\calp}{{\cal P}}
\newcommand{\calk}{{\cal K}}
\newcommand{\calr}{{\cal R}}
\newcommand{\calra}{{\cal R}_{{\cal A}}}
\newcommand{\calrn}{{\cal R}_{0}}
\newcommand{\cals}{{\cal S}}

\newcommand{\calv}{{\cal V}}
\newcommand{\calvc}{{\cal V}_{cons.}}
\newcommand{\calz}{{\cal Z}}
\newcommand{\eeq}{\end{equation}}
\newcommand{\ellad}{\ell_{ad}}
\newcommand{\ellf}{\ell_N}
\newcommand{\ellt}{\ell_3}
\newcommand{\ellv}{\ell}
\newcommand{\elln}{\ell_0}
\newcommand{\ellnp}{\ell_0^+}
\newcommand{\real}{{\rm Re}}
\newcommand{\tr}{{\rm tr}\;}
\newcommand{\vcl}{{\cal V}}
\newcommand{\veff}{{\cal V}_{e\!f\!f}}
\newcommand{\veig}{{\cal V}_{V\!dm}}
\newcommand{\vmf}{{\cal V}_{mf} }
\newcommand{\vx}{\vec{x}}
\newcommand{\zn}{$Z(N)\;$}

\def\cmp#1#2#3{Comm. Math. Phys. {\bf #1}, #2 (#3)}

\def\ibid#1#2#3{{\it ibid.} {\bf #1}, #2 (#3)}
\def\ijm#1#2#3{Intl. Jour. Mod. Phys. {\bf #1}, #2 (#3)}
\def\jhep#1#2#3{Jour. High Energy Phys. {\bf #1}, #2 (#3)}

\def\npb#1#2#3{Nucl. Phys. B {\bf #1}, #2 (#3)}
\def\npsb#1#2#3{Nucl. Phys. Proc. Suppl. B {\bf #1}, #2 (#3)}
\def\plb#1#2#3{Phys. Lett. B {\bf #1}, #2 (#3)}
\def\prc#1#2#3{Phys. Rev. C {\bf #1}, #2 (#3)}
\def\prd#1#2#3{Phys. Rev. D {\bf #1}, #2 (#3)}
\def\prl#1#2#3{Phys. Rev. Lett. {\bf #1}, #2 (#3)}
\def\phr#1#2#3{Phys. Rep. {\bf #1}, #2 (#3)}

\def\rpp#1#2#3{Rep. Prog. Phys. {\bf #1}, #2 (#3)}
\def\rmp#1#2#3{Rev. Mod. Phys. {\bf #1}, #2 (#3)}
\def\sjnp#1#2#3{Sov. Jour. Nucl. Phys. {\bf #1}, #2 (#3)}
\def\zpc#1#2#3{Z. Phys. C {\bf #1}, #2 (#3)}
\begin{document}
\title{Deconfinement in Matrix Models about the Gross--Witten Point}
\author{Adrian Dumitru,$^{a}$
Jonathan Lenaghan,$^b$ and Robert D. Pisarski$^{c}$}
\affiliation{
$^a$Institut f\"ur Theoretische Physik, J.~W.~Goethe Univ.,
Postfach 11 19 32, 60054 Frankfurt, Germany\\
$^b$Dept. of Physics, Univ. of Virginia, Charlottesville, VA, 22904, U.S.A.\\
$^c$High Energy Theory \& Nuclear Theory Groups, 
Brookhaven National Lab., Upton, NY, 11973, U.S.A.;\\
Niels Bohr Institute, Blegdamsvej 17, 2100 Copenhagen, Denmark;\\
Frankfurt Institute for Advanced Study, 
J.~W.~Goethe Univ., Robert Meyer Str. 10, D-60054 Frankfurt, Germany\\
}
\begin{abstract}
We study the deconfining phase transition
in $SU(N)$ gauge theories at nonzero temperature
using a matrix model of 
Polyakov loops. The most general effective action, including all
terms up to two spatial derivatives, is presented.
At large $N$, the action is dominated by the loop potential:
following Aharony {\it et al.}, we show how 
the Gross--Witten model represents an ultra-critical point
in this potential.
Although masses vanish at the Gross--Witten point,
the transition is of first order, as the fundamental loop jumps
only halfway to its perturbative value.
Comparing numerical analysis of the $N=3$ matrix model to lattice simulations,
for three colors the deconfining transition appears to be near
the Gross--Witten point.  To see if this persists
for $N \geq 4$, we suggest measuring
within a window $\sim 1/N^2$ of the transition temperature.
\end{abstract}
\date{\today}
\maketitle

\section{Introduction}

As an $SU(N)$ gauge theory is heated, it undergoes
a phase transition from a confined, to a deconfined, phase.
The standard order parameter for deconfinement
is the Polyakov loop in the fundamental representation \cite{old_loop}.
On the lattice, numerical simulations universally measure 
the expectation value of this loop to see where the deconfining
transition occurs \cite{lattice_review}.  
This is only of use at finite lattice spacing, though,
since in the continuum limit, it appears that
the expectation values of all bare loops vanish \cite{dhlop}.

Recently, two techniques have been developed to measure renormalized
loops, which are nonzero
in the continuum limit \cite{dhlop,bielefeld}.
A Polyakov loop represents the propagation of an infinitely massive,
``test'' quark.  On a lattice, even an infinitely massive field 
undergoes an additive mass shift, which diverges 
in the continuum limit.  This mass shift 
generates a renormalization constant for the loop:
as for any bare quantity, this must be divided out in
order to obtain the expectation value of the renormalized loop.

Following Philipsen \cite{philipsen}, one method computes
the singlet potential \cite{bielefeld}, which is equivalent
to computing the Wilson loop at nonzero temperature.
When the Wilson loop is narrow in the spatial direction, it
can be computed in perturbation theory, allowing for the
renormalization constant of the loop to be extracted
\cite{cusp,difference}.

The second method involves the direct measurement of one point
functions of Polyakov loops.  The drawback is that
these must be measured on several lattices:
all at the same physical temperature, but with different
values of the lattice spacing \cite{dhlop}.
While these two methods are rather distinct, for 
the triplet loop in a $SU(3)$ gauge theory, 
they agree to within $\approx 10\%$ over temperatures from
$T_d \rightarrow 3 T_d$, where $T_d$ is the transition temperature
for deconfinement \cite{cusp}.

In an asymptotically free theory, the expectation value of a
Polaykov loop approaches one at high temperature.  It is perfectly
conceivable that when a gauge theory deconfines, that it goes
directly to a gluon plasma which is close to perturbative, even
at temperatures just above $T_d$.  
(An example of this is
Fig. (\ref{fig:first}) of Sec.~\ref{away_gross_witten}.)
Indeed, if the deconfining
phase transition is strongly first order, this is what one might
naively expect.  From lattice simulations, while the deconfining
transition is of second order for two colors \cite{two_colors_A,two_colors_B},
it is of first
order for three colors \cite{lattice_review,three}, 
and becomes more strongly first order
as the number of colors increases.  
Results for four \cite{four_colors,teper,meyer}, six, and eight
colors \cite{teper,meyer} suggest that at large $N$,
the latent heat is proportional to an obvious factor of $\sim N^2$,
from the overall number of gluons in a deconfined phase, times a constant.

For three colors, the lattice measurements of 
the renormalized triplet loop are admittedly preliminary.
With that caveat, the expectation value of the renormalized 
triplet loop is found to be near one, $\approx .9 \pm 10\%$,
for temperatures which are 
as low at $\approx 3 T_d$.  This agrees with resummations of
perturbation theory, which work down from infinite temperature,
but consistently fail at a temperature which is something like
$\approx 3 \rightarrow 5 T_d$ \cite{resum}.

Just above the transition, however, the renormalized triplet
loop is far from one: it is only $\approx 0.4 \pm 10\%$ at $T_d^+$.
This suggests the following picture.
For temperatures above $\approx 3 T_d$, 
the deconfined plasma is, after resummation, nearly
perturbative.  For temperatures between $T_d$ and $\approx 3 T_d$,
however, the theory is in a 
phase which is deconfined but {\it non}-perturbative.  
The existence of such a non-perturbative (quark) gluon plasma may be
indicated by the experimental data on heavy ion collisions
at RHIC~\cite{RHIC}.

At large $N$, the deconfining transition has been studied
using string and duality methods 
\cite{hagedorn,previous,aharony,fss,schnitzer,aharony2}.
At infinite $N$, by 
restricting the theory to live on a space-time where space is a very
small sphere, it is possible to compute the Hagedorn temperature
analytically 
\cite{hagedorn,previous,aharony,fss,schnitzer,aharony2}.  
In this limit, Aharony {\it et al.} \cite{aharony} catagorized
the possible relationships between
the deconfining, and the Hagedorn, temperatures.

In this paper we develop a general approach to the deconfining
transition, based upon a matrix model of Polyakov loops.
We begin in Sec.~\ref{sec_EffModels}
with the most general effective Lagrangian,
including all terms up to two spatial derivatives.
At first sight, matrix models of loops appear to
be an inelegant type of nonlinear sigma model
\cite{banks_ukawa,loop1,loop2a,loop2b,loop2c,loop2d,loop2e,loop2f,loop3,sannino,ogilvie,interface1,interface2}.
This is not a matter of choice, but
is dictated by the physics.  For ordinary
nonlinear sigma models, in the chiral limit there are no
potential terms, only terms involving spatial derivatives.
These can be studied in a mean field approximation, but this is
only valid when the number of spacetime dimensions is large
\cite{zinn,drouffe_zuber,largeN}.
In contrast, the effective action for
loops starts with a potential, independent of 
spatial derivatives.  Because of this, 
a systematic large $N$ expansion can be developed
in any number of spacetime dimensions.  

The large $N$ expansion of the loop potential is developed in
Sec.~\ref{sec_InfiniteN}.  
We show how the Vandermonde determinant, which appears
in the measure of the matrix model, gives a type of
potential term \cite{brezin,gross_witten,aharony}.
also contributes a potential term.
At infinite $N$, the vacua of the theory are then given
by the stationary points of an effective potential, which is
the sum of the loop, and what we call the Vandermonde, potentials.

In the loop potential, the adjoint loop is a mass term, while
loops in higher representations represent interaction terms.
We show how the Gross--Witten model \cite{gross_witten,jurk}, which
arose in a very different context, 
corresponds to a loop potential which is purely a mass
term, with no higher terms.  We term the point at which the
deconfining transition occurs 
as the ``Gross--Witten point'' \cite{dhlop}.  At this point, the
expectation value for the 
loop in the fundamental representation, $\ell$, jumps from $0$
to precisely $\frac{1}{2}$ 
\cite{dhlop,aharony,gross_witten,kogut}.
The loop potential is most unusual, however: it vanishes
identically for $\ell$ between $0$ and 
$\frac{1}{2}$, and is only nonzero when 
$\ell \geq \frac{1}{2}$ (see Fig.~\ref{fig:gw} below).  
This is only possible because of the contribution of the Vandermonde
potential.  This potential, which is not analytic in $\ell$
at $\ell = \frac{1}{2}$, produces a deconfining transition which
is thermodynamically of first order \cite{dhlop,aharony,kogut}.
This flat potential also implies that at the transition point,
masses vanish, asymmetrically, in both phases 
(these are the masses
for the fundamental loop, along the direction of the condensate)
\cite{dhlop}.  
If a background field is added, at the Gross--Witten point
the first order transition turns into one of third order, no matter how
small the background field is \cite{green_karsch,damgaard_patkos}.

Aharony {\it et al.} analyzed the transition when
quartic interactions are included 
\cite{aharony}.  They suggested that in the space
of loop potentials, the Gross--Witten model
represents a tri-critical point.
We refine their analysis, and extend it to include arbitrary interactions of
the fundamental loop.  We discuss why 
the Gross--Witten point is more accurately described as an
ultra-critical point, where all higher interactions vanish.
Physically, this is defined as 
the unique point where the transition is of first
order, and yet masses vanish; it only arises at infinite $N$.

We analyze the matrix integral by introducing a constraint field
for the fundamental loop \cite{zinn}.  
Various constraint fields can then be integrated out in different
order.  One order is equivalent to computing the Vandermonde potential
by Legendre transformation 
\cite{aharony,fss,schnitzer,aharony2}.
We show that the other order is equivalent to what was called,
previously, a mean field approximation 
\cite{kogut,green_karsch,damgaard_patkos,dhlop}.
If the potential includes just the adjoint loop,
we check that while these two methods give different effective
potentials, that at any stationary point, 
both the expectation values of the loop, and masses, agree.

For three colors, 
the surprising aspect of the lattice data is that the value
of the renormalized triplet loop at $T_d^+$, 
$\approx 0.4$, is close to $\frac{1}{2}$.  Further, masses associated with
the triplet loop --- especially the string tension --- decrease
significantly near the transition \cite{three}.  This suggests that the
transition for three colors is close to the Gross--Witten point at
infinite $N$.

The lattice results provide further evidence for the utility of
using a large $N$ expansion at $N=3$.  
The expectation values of 
renormalized sextet and octet loops, which are expected to
be $\sim 1/N$ or smaller, never exceed
$\approx 25\%$ at any temperature, and are usually much less \cite{dhlop}.

In Sec.~\ref{sec_finiteN} we study the loop potential for finite $N$.
Following Damgaard {\it et al.}~\cite{damgaard}, 
we start by showing that when fluctuations
in the matrix model are neglected, 
that the expectation value of {\it any} loop --- including those
which are \zn neutral --- vanish in the confined phase.
This is a striking difference between matrix models and more
general \zn symmetric theories \cite{kiskis}.

The matrix model is merely a two dimensional integral, which
can be studied numerically.  In Sec.~\ref{sec_finiteN}
we investigate theories close to the $N=3$ analogy of the
Gross--Witten point, including cubic interactions.
The matrix model gives the expectation values of
loops in arbitrary representations.  
We also compute the mass of the loop, which is
related to the (gauge invariant) Debye mass, as obtained
from the two point function of Polyakov loops.  
We discuss how measurements of both the Debye mass, and the
renormalized triplet loop, help probe the loop potential.

We also use the matrix model to
compute difference loops \cite{dhlop} for the sextet,
octet, and decuplet loops.  If only the loop potential is
included, the difference loops from lattice simulations \cite{dhlop}
are not well reproduced.  
Presumably it is necessary to include fluctuations, due to kinetic
terms, in the effective theory.
Nevertheless, up to these small corrections, $\leq 20\%$ at all
temperatures, it is clear that the deconfining transition for
three colors is close to the Gross--Witten point at infinite $N$.

In the Conclusions, Sec.~\ref{sec_Summary}, we discuss why lattice
data \cite{four_colors,teper,meyer} may indicate that
the deconfining
transitions for $N\geq 4$ are also near the Gross--Witten point.
We suggest that one must be {\it very} 
close to the transition to see this, 
when the reduced temperature $|T-T_d|/T_d\sim 1/N^2$.

\section{Effective Models} 
\label{sec_EffModels}

\subsection{Nonlinear Sigma Models}
\label{sigma}

We begin by briefly reviewing effective theories for nonlinear sigma models
\cite{zinn}.  For definiteness, consider the model appropriate
to chiral symmetry breaking for $N_f$ flavors, 
with a global symmetry group of \cite{donoghue}
\beq
G_f = SU_L(N_f) \times SU_R(N_f) \; ,
\eeq
and a field $U$, transforming as 
\beq
U \rightarrow e^{2 \pi i j/N_f} \; \Omega_L^\dagger \; U \; \Omega_R \; ;
\eeq
where $\Omega_{L,R}$ are $SU_{L,R}(N_f)$ transformations.
Because the 
left and right handed chiral rotations are distinct, in the
chiral limit, terms such as the trace of $U$, $\tr U$, cannot arise.
Thus in the chiral limit, there is no potential for $U$, and 
there are only derivative terms.  These start at second order:
\beq
{\cal L} = f^2_\pi \; \tr |\partial_\mu U|^2 + \ldots \; ;
\label{sigma_kinetic}
\eeq
subject to the constraint that $U$ is a unitary matrix,
\beq
U^\dagger U = {\bf 1}_{N_f} \; .
\label{sigma_constraint}
\eeq
The series then continues with terms of quartic order in derivatives 
\cite{donoghue}.

After chiral symmetry breaking, what remains is a vector symmetry
of $SU_V(N_f)$, with $\Omega_L = \Omega_R = \Omega_V$.  Then there
is a potential for $U$ possible,
\beq
\vcl = m^2_\pi \; \tr U \; + \ldots .
\label{sigma_potential}
\eeq
This potential is manifestly proportional to the pion mass squared,
since it must vanish in the chiral limit.  
Near the chiral limit, 
the kinetic terms in (\ref{sigma_kinetic}) dominate over the
potential term in (\ref{sigma_potential}) if the volume of
space (or space-time) is large.  If the volume is small, however,
the potential term can dominate, as the modes with nonzero
momentum are frozen out \cite{epsilon}.

Other sigma models are constructed by allowing the transformations
$\Omega_L$ and $\Omega_R$ to be equal.  For example, take $U$ to be
a $SU(2 N_f)$ matrix, and impose a further constraint, such as that
the trace of 
$U$ vanishes, $\tr \, U = 0$.  In this case, the
symmetry is that of a symmetric space,
$G_f = SU(2 N_f)/S(U(N_f) \times U(N_f))$ \cite{zinn,symmetric}.
Because the trace of $U$  is constrained to be a fixed number,
there is no potential possible, and the action is given
entirely by kinetic terms, as in (\ref{sigma_kinetic}).  
If the trace is constrained to be some other value, then
the symmetry group changes, but there is still no potential
possible.

These nonlinear sigma models are 
renormalizable in an expansion about two space-time
dimensions.  There is a phase with broken symmetry above
two dimensions, where the expectation value of $U$ is
nonzero.  This expectation value is generally not
proportional to the unit matrix, and there are Goldstone bosons
in the broken phase.  In the chiral limit, as there is no potential for $U$,
it is not easy studying the possible patterns of symmetry
breaking directly in the nonlinear form of the model
(linear models are usually more useful).  Symmetry
breaking can be studied using mean field theory \cite{zinn,drouffe_zuber}.  
To do so, the continuum form of the theory is replaced
by a lattice form, with matrices $U$ on each site $i$.
In that case, the next to nearest neighbor interaction at
a site $i$, for unit lattice vector $\hat{n}$, becomes,
in mean field approximation, 
\beq
\tr \left( U_i U_{i + \hat{n}}\right) 
\rightarrow \tr \left( U_i \; \langle U \rangle \right) \; .
\eeq
In this case, mean field theory is only applicable when
the number of nearest neighbors, or space-time dimensions, 
is very large \cite{zinn,drouffe_zuber}.

\subsection{Polyakov Loops: Preliminaries}

We discuss the deconfining transition at a nonzero temperature $T$,
for a space of infinite volume.  The analysis is
more general, though,
and can be easily extended to the case in which
space is of finite extent, {\it etc.}

The thermal Wilson line is
\beq
\boldl_\calr(\vx) = \calp \; \exp 
\left(  i g \int^{1/T}_0 A_0^a(\vx,\tau) \;
{\bf t}^a_\calr \; d \tau \right)  \; .
\eeq
We follow our previous notations and conventions \cite{dhlop}.
We define the thermal Wilson line at a point $\vx$ in space,
letting it run all of the 
way around in imaginary time, from $0$ to $\tau = 1/T$.
Otherwise, $\calp$ denotes path ordering, 
$g$ is the gauge coupling constant, 
and $A_0^a$ the vector potential in the time direction.
The ${\bf t}^a_\calr$ are the generators of $SU(N)$ in 
a representation $\calr$, which is taken to be irreducible.

The Wilson line transforms homogeneously 
under local gauge transformations, 
\beq
\boldl_\calr(\vx)
\rightarrow \Omega_\calr^\dagger(\vx,1/T) \; \boldl_\calr(\vx)\;
\Omega_\calr(\vx,0) \; .
\eeq
For gauge transformations which are periodic in imaginary time,
we form a gauge invariant quantity, the Polyakov loop in $\calr$,
by taking the trace,
\beq
\ell_\calr = \frac{1}{d_\calr} \; \tr \, \boldl_\calr \; .
\eeq
We define the Polyakov
loop as the normalized trace of $\boldl_\calr$,
with $d_\calr$ equal to the dimension of $\calr$.  The advantage
of this is that in an asymptotically free theory, 
all Polyakov loops are of unit magnitude in the limit of
infinitely high temperature.
We denote representations by their dimensionality,
with two exceptions, which differ from previous use \cite{dhlop}.
The Wilson line in the fundamental representation is $\boldl$,
with the fundamental loop
\beq
\ellf = \frac{1}{N} \;  \tr \, \boldl \; .
\eeq
The adjoint loop is denoted
\beq
\ellad = \frac{1}{N^2 - 1} \left( \, |\tr \boldl|^2 - 1 \right) \; .
\eeq

Besides local gauge transformations, which are strictly periodic
in time, in a pure gauge theory there are also 
global gauge transformations,
which are only periodic up to an element of the center of the gauge
group.  For an $SU(N)$ gauge group, the center is
\zn.  In the fundamental representation, the
simplest global \zn transformation is
\beq
\Omega(\vx,1/T) = 
e^{2 \pi i/N} \; \Omega(\vx,0)  \; .
\label{phase_1}
\eeq
Defining the charge of the fundamental
representation to be one, the charge of an arbitrary representation,
$e_\calr$, follows from its transformation under (\ref{phase_1}),
\beq
\ell_\calr \rightarrow e^{2 \pi i e_\calr /N} \ell_\calr \; .
\eeq
As a cyclic group, the charge $e_\calr$ is only defined modulo $N$.
A special set of representations are those with zero \zn charge,
which we denote as $\calrn$.  The simplest example is the adjoint
representation.

Below the deconfining transition temperature $T_d$, the expectation
values of all loops with nonzero \zn charge vanish,
\beq
\langle \ell_\calr \rangle = 0 \;\;\; , \;\;\;
T \leq T_d \;\;\; , \;\;\; \calr \neq \calrn \; ,
\eeq
and the theory is in a \zn symmetric phase.

The global \zn symmetry is broken in
the deconfined phase, as loops in all representations condense,
\beq
\langle \ell_\calr \rangle \neq 0 \;\;\; , \;\;\;
T \geq T_d \;\;\; , \;\;\; \forall \; \calr \; .
\eeq

In the limit of large $N$, factorization \cite{dhlop,drouffe_zuber,largeN}
implies that all expectation
values are powers of that for the fundamental, and anti-fundamental, loops:
\beq
\langle \ell_R \rangle = 
\langle \ellf \, \rangle^{p_+} \; 
\langle \ellf^* \rangle^{p_-} \;\;\; , \;\;\; N = \infty \; .
\label{factorization}
\eeq
The integers $p_+$ and $p_-$ are defined in \cite{dhlop}.
For the adjoint, for example,
$p_+ = p_- = 1$; in general, the \zn charge $e_\calr = p_+ - p_-$.
Factorization is exact only at infinite $N$,
but holds for all temperatures.
Corrections to factorization are
of order $\sim 1/N$ at large $N$; for the adjoint, they are
$\sim 1/N^2$ \cite{dhlop}.

In general, loops with zero \zn charge, such as the adjoint,
can have nonzero expectation values at all temperatures, including
in the confined phase.  At infinite $N$, factorization implies
that they vanish below $T_d$, because the expectation value of the
fundamental loop does.  An important question to which we shall
return is the extent to which \zn neutral loops condense in the
confined phase.

\subsection{Matrix Models of Polyakov Loops}
\label{matrix_models_loops}

In the same spirit as familiar for effective models of chiral
symmetry breaking \cite{donoghue,epsilon}, discussed in Sec. \ref{sigma},
we construct an effective theory for deconfinement, applicable
over distances greater than the inverse temperature.  Thus
fields depend only upon the spatial coordinates, $\vec{x}$.
We take as the basic field the Wilson line, $\boldl(\vec{x})$;
as that is a gauge covariant field, we need also include
the spatial components of the vector potential, $A_i(\vec{x})$.

We are then led to construct the most general action consonant
with the relevant symmetries, which are those of
global {\zn}, and local $SU(N)$, transformations.  We will
allow the vacuum to spontaneously break the \zn symmetry,
but do not allow $SU(N)$ to break.  (It
would be interesting to extend the following analysis to include a
Higgs phase, as is appropriate for symmetry restoration in 
the electroweak interactions.)  

We start with potential terms, independent of any derivatives
in space-time.  Because of the left-right symmetry, for sigma models
there is no potential in the chiral limit.  In constrast,
loop models always have a potential.
To be invariant under $SU(N)$ transformations, terms in
the potential must be a sum of loops; to be invariant
under \zn, these must be \zn neutral:
\beq
\vcl(\boldl) = - m^2 \, \ellad + \sum_{\cals \epsilon \calrn} 
\; \kappa_\cals \; \real \; \ell_\cals \; ;
\label{general_potential}
\eeq
Loops can be complex valued when $N \geq 3$,
so in the potential we take the real part of each loop.

By using group theory, this potential can be rewritten in many
different ways.  
As higher powers of any loop can always be reexpressed as a linear
sum over loops in other representations, this potential is
the most general form possible.  

In $\vcl(\boldl)$, the adjoint loop looks like a mass term for the
fundamental loop, while loops in higher representations
look like interactions of the fundamental loop (plus new
terms, such as $\tr \boldl^2/N$, {\it etc.}).  
The loop potential differs from that of ordinary scalar fields,
though.  In the potential of a scalar field, the coupling constant
for the highest power of the field must be positive
in order for a ground state to exist.
This isn't true for the loop potential, as each loop is bounded
by one: there is no constraint
whatsoever on the signs of the mass squared, nor on any of the coupling
constants.  For example, we choose the sign of the adjoint loop
to have a negative sign (so that $m^2 \sim T-T_d$, as for ordinary
spin systems).  Then $m^2 \rightarrow - \infty$ drives one to
the confined phase, while $m^2 \rightarrow + \infty$
drives one into the deconfined phase.  Further, it is consistent
to have just a mass term, with no other terms in the potential,
setting all $\kappa_{\cal S} = 0$.  Thus 
while the adjoint loop looks like
a mass term for the fundamental loop, because the basic variable
is $\boldl$, and not $\ell_N$, the theory is still nontrivial
with no ``interactions'', $\kappa_{\cal S} = 0$.

At each point in space, the Wilson line can be diagonalized
by a local gauge transformation, and so depends upon $N-1$
eigenvalues.  Thus instead of $\boldl$, it is also possible
to chose a set of $N-1$ loops, and rewrite the loop potential
in terms of these.  We find this of use in Sec. IV, 
(\ref{potential_sum_calra}). 

The loop potential can be computed analytically in two limits:
when the temperature is very high \cite{interface1,interface2},
and when space is a small sphere
\cite{hagedorn,previous,aharony,fss,schnitzer,aharony2}.  
For both cases there is a large mass scale, either the temperature,
or the inverse radius of the sphere, so that the effective
gauge coupling is small by asymptotic freedom.
In perturbation theory, 
the loop potential is computed from the one loop determinant in
the presence of a background gauge potential $A_0 \sim \log{\boldl}$
\cite{hagedorn,previous,aharony,fss,schnitzer,aharony2,interface1,interface2}. 
At very high temperature (and infinite spatial volume),
while the result can be 
written in terms of an infinite polynomial of (traces of) powers of $\boldl$,
as in (\ref{general_potential}), the sum can be explicitly performed.
The result is just a a simple quartic potential 
of $A_0$ \cite{interface1}.

In a perturbative regime, it is natural that the loop
potential is more transparent in terms of elements of the Lie
algebra, the $A_0$, instead of elements of the Lie group, the $\boldl$.
This suggests that an effective theory of loops is not especially
convenient when the theory is essentially perturbative.
For $SU(3)$, resummations of perturbation theory suggest that 
the perturbative regime appears for
$T \approx 3 T_d$ \cite{resum}.  Thus for $SU(3)$, loops are only useful
for temperatures below $\approx 3 T_d$.

The loop potential can also be computed on a small sphere
\cite{previous,aharony,fss,schnitzer,aharony2}.  
There are now two scales in the problem --- the radius of
the system, and the temperature --- so the 
result is much more complicated than for infinite volume.
The result is a sum like that of (\ref{general_potential}).

Kinetic terms, involving two derivatives, are constructed similarly.
While sigma models only have one term with two derivatives, 
(\ref{sigma_kinetic}), due to the change
in symmetry, loop models have an abundance.
One class of terms involves covariant derivatives of the Wilson line:
\beq
\calk_1(\boldl) = \frac{1}{\widetilde{g}^2} \; \tr |D_i \boldl|^2 
\left(
1 + \sum_{\cals \epsilon \calrn} \; \lambda_\cals\;
\; \real \, \ell_\cals \right)
\label{electric_loop}
\eeq
The first term is a sort of ``electric loop'':
it is the original electric field of the gauge theory, rewritten
in terms of loops.  
At tree level, the coupling constant $\widetilde{g}$ equals the
gauge coupling $g$, but in general, it is an independent coupling
constant of the effective theory.  Besides the electric loop,
there is an infinite series 
of terms involving loops in \zn neutral representations.
The couplings of magnetic fields is similar to (\ref{electric_loop}):
there is an infinite series of \zn neutral loops which couple to
the (trace of the) magnetic field squared.  

There are also derivative terms for the loops themselves:
\beq
\calk_2(\boldl) = 
\frac{1}{\widetilde{g}^4} \; \sum_{\cals,\cals',\cals'' \epsilon \calr}
\; \zeta_{\cals,\cals',\cals''}
\;  \real \; \left( \partial_i \ell_\cals \right) 
\left( \partial_i \ell_{\cals'} \right) \; \ell_{\cals''}
\; ,
\eeq
where the representations are constrained so that the total
\zn charge of each term vanishes, 
$e_{\cals} + e_{\cals'} + e_{\cals''} = 0$, modulo $N$.

Lastly, there are terms involving two derivatives of $\boldl$,
coupled to a field with \zn charge minus two:
\beq
\calk_3(\boldl) = \frac{1}{\widetilde{g}^2} 
\; \tr \; \real \; (D_i \boldl)^2 \; 
\sum_{\cals \epsilon \calr}
\; \widetilde{\zeta}_\cals  \; \ell_\cals \;\;\; , \;\;\; 
e_\cals = -2 \; .
\eeq
For three colors, this series starts with the triplet loop.

In the deconfined phase, the free energy is of order $\sim N^2$.
Since all loops are of order one at large $N$, 
in the action we always assume that the potential
$\vcl$ is multiplied by an overall factor of $\sim N^2$
at large $N$.  The mass $m^2$ is then naturally
of order one at large $N$; how the coupling constants $\kappa_\cals$
scale with $N$ is discussed in Sec. \ref{couplings_largeN}.
For the kinetic terms, to contribute $\sim N^2$, the coupling 
$\widetilde{g}$ should scale like the gauge coupling,
with $g^2 N$ and $\widetilde{g}^2 N$ fixed at large $N$;
then the couplings 
$\zeta_{\cals,\cals',\cals''}$ and $\widetilde{\zeta}_\cals$
are both $\sim 1$.

The renormalization of the loop potential was discussed in
\cite{dhlop}.  The Wilson line in a given, irreducible representation
undergoes mass renormalization, $m^{div}_\calr$; on a lattice,
this is a power series in the coupling constant times the lattice
spacing, $1/a$.
For a Wilson line at 
a temperature $T$, the renormalization constant, $\calz_\calr$,
and the renormalized loop, $\widetilde{\ell}_\calr$, are given by
\beq
\widetilde{\ell}_\calr = \ell_\calr/\calz_\calr \;\;\; , \;\;\
\calz_\calr = \exp(-m^{div}_\calr/T) \; .
\label{mass_ren}
\eeq
Loops in irreducible
representations do not mix under renormalization.

The renormalization constants for the kinetic terms are similar,
but more involved, than those for the potential.
All kinetic terms undergo mass renormalization, with
renormalization constants which are the exponential of a
divergent mass times the length of the loop.
For loops without cusps, there is no condition to fix the
value of these renormalization constants at some scale, while
loops with cusps do require such a condition \cite{dhlop,cusp,difference}. 
For kinetic terms, additional renormalization constants, and
conditions to fix their value at some scale, may be required.
For example, each term in (\ref{electric_loop}) experiences
mass renormalization.  In addition, since 
the term $\sim 1/\widetilde{g}^2$ arises from the electric field in the
bare action, at least in four spacetime dimensions,
it will require an additional renormalization constant related
to coupling constant renormalization.

We have written down all possible terms involving two derivatives,
but this classification may well be overly complete.  
In particular, renormalization will greatly restrict the possible
terms.  The entire list is relevant in 
two spatial dimensions; the associated $\beta$-functions
can be computed in the ultraviolet limit \cite{beta}.
In three spatial dimensions, 
a scalar field has dimensions of the square
root of mass, and most of the above terms are non-renormalizable,
and so can be ignored.  

Terms which arise from the coupling
to fields which are not \zn invariant, such as quark fields,
can also be included.
The simplest possible coupling involves the trace of the chiral
field:
\beq
m^2_\pi \; \tr U \; \real \; \ellf \; .
\eeq
This, however, is chirally suppressed, proportional to
$m_\pi^2$.  There are couplings which are not
chirally suppressed, but these necessarily involve derivatives
of the chiral field; these start as
\beq
\tr |\partial_\mu U|^2 \; \real \; \ellf \; .
\eeq
We do not know if this chiral suppression significantly affects
the breaking of \zn symmetry; in mean field theory, this can
be analyzed by using a nonlinear sigma model for the chiral fields \cite{mean}.
For an analysis of \zn breaking terms 
in terms of linear sigma models, see 
Moscy, Sannino, and Tuominen \cite{sannino}.

\section{Infinite N} 
\label{sec_InfiniteN}

\subsection{Gross--Witten Point}
\label{gross_witten_point}

The effective action for loops is much more involved than for sigma
models.  Because loops have a potential, however, 
we can systematically perform a large $N$ expansion by 
minimizing an effective potential.  Corrections in $1/N$ arise
from fluctuations, which arise when spatial derivatives are included.
In the rest of the paper, we ignore fluctuations, and the 
kinetic terms of Sec.~\ref{matrix_models_loops}, to concentrate on 
what is a matrix-valued mean field theory.  
Sec. III treats this matrix model at infinite $N$; Sec. IV, finite $N$.
Even at infinite $N$, we stress that 
because we cannot compute the loop potential from first principles,
all we are doing is characterizing all possible transitions
in terms of couplings in the loop potential, the $\kappa_\cals$
of (\ref{general_potential}).

If fluctuations are neglected, we can assume that the Wilson line
is the same at each point in space, $\boldl(\vec{x}) = \boldl$.  Thus
the functional integral reduces to a single integral,
\beq
\calz = \int d\boldl \; \exp\left( - N^2 \vcl(\boldl) \right) \; .
\label{partition_function}
\eeq
At large $N$, factorization implies that for arbitrary
representations, normalized loops 
are just products of the fundamental, and anti-fundamental, loop \cite{dhlop}.
Then the loop potential 
reduces simply to a power series in $|\ellf|^2$,
\beq
\vcl(\boldl) 
= - m^2 |\ellf|^2 + \kappa_4 \left( |\ellf|^2\right)^2
+ \kappa_6 \left( |\ellf|^2\right)^3 + \ldots \; .
\label{class_pot}
\eeq
We have relabeled the couplings $\kappa_{\cal S}$ in 
(\ref{general_potential}) as 
$\kappa_{2n}$, where the subscript now denotes the power 
of $\ellf$.  The $U(1)$ symmetry
is broken to \zn by a term $\sim \kappa_N \, \real \, (\ell_N)^N$, but this is
negligible at large $N$.   

As the potential is multiplied by an overall factor
of $\sim N^2$, and the loops are normalized to be of order one,
the natural guess is also to take the 
couplings $\kappa_{\cal S}$ of order one at large $N$.
As we discuss in Sec. \ref{couplings_largeN},
this choice of couplings is far from innocuous.
The advantage is that this assumption vastly simplifies the
analysis, and allows us to gain insight which is otherwise
obscured by technical complications \cite{aharony}.
Further, as far as the bulk thermodynamics
is concerned, the only novel behavior emerges in this regime; otherwise,
the deconfining transition is rather ordinary.

To find the stationary points of the functional integral, we
introduce a delta-function into the integral \cite{zinn} as
\beq
\calz = \int d\boldl \; \int d\lambda \;\;
\delta(\lambda - \ellf) \; \exp\left( - N^2 \vcl(\lambda) \right) \; .
\label{partition_function_constrained}
\eeq
Here $\lambda$ is a (complex) number, equal to the value of 
$\ellf = \tr \boldl/N$ for a given matrix $\boldl$.  
As such, in the action we can replace $\vcl(\ellf)$ by $\vcl(\lambda)$.
The constraint is then exponentiated by introducing a field 
$\overline{\omega}$,
$$
\calz = \int d\lambda \int d\overline{\omega} \int d\boldl \;
\exp\left( - N^2 \calvc \right) \; ,
$$
\beq
\calvc = 
\vcl(\lambda) + i \overline{\omega} (\lambda  - \ellf) \; .
\label{constrained_pot}
\eeq
In this expression, both $\lambda$ and $\overline{\omega}$
are complex numbers, not matrices, or traces thereof.
Usually, $i \overline{\omega}$ is real at a stationary point,
and so we define $\omega = i \overline{\omega}$.

The full integral is over 
$\lambda$, $\overline{\omega}$, and $\boldl$.  We break off the
integral over $\boldl$, and define that piece as
\beq
\calz_{GW}(\omega) = \int d\boldl \; \exp\left( N^2 \, \omega \; 
\real \, \ellf \right) \; .
\label{gross_witten_integral}
\eeq
By an overall \zn rotation, we can choose any condensate for $\ellf$
to be real and positive.  Both $\lambda$ and $\omega$ are complex
fields, but to look for a real stationary point in $\ellf$, we need
only consider the real parts. 

The integral $\calz_{GW}$ was done by 
Brezin {\it et al.} \cite{brezin}, Gross and Witten \cite{gross_witten},
and Aharony {\it et al.} \cite{aharony}.
For the unitary matrix $\boldl$, 
all that matters are the eigenvalues of $\boldl$, $\boldl_{i,j}
= \delta_{i j} \exp(i \alpha_i)$.  
In the limit of infinite $N$, the number of eigenvalues is infinite,
and it is convenient to introduce the density 
of eigenvalues, $\rho(\alpha)$.  The solution for this density is:
\beq
\rho(\alpha) = \frac{1}{2 \pi} \left( 1 + \omega \, \cos \alpha \right)
\;\;\; , \;\;\; \omega \leq 1 \; ;
\eeq
\beq
\rho(\alpha) = \frac{1}{\pi} \; \cos \frac{\alpha}{2}
\left( 1 - \omega \; \sin^2\frac{\alpha}{2} \right)^{1/2} 
\;\;\; , \;\;\; \omega \geq 1 \;
\eeq
where the latter only holds for $\omega \; \sin^2(\alpha/2) < 1$.

We now make the following observation.  The value of the condensate is
\beq
\ell_0 = \int^\pi_{- \pi} d\alpha \; \rho(\alpha) \cos \alpha \; .
\eeq
With these expressions for the eigenvalue density, it is not
difficult to show that 
\beq
\ell_0(\omega) = \frac{\omega}{2} \;\;\; , \;\;\; \omega \leq 1 \; ;
\label{elln_omA}
\eeq
and
\beq
\ell_0(\omega) = 1 - \frac{1}{2\omega} \;\;\; , \;\;\; \omega \geq 1 \; ;
\label{elln_omB}
\eeq

To implement these relations between the expectation value and $\omega$,
we introduce yet another constraint field, $\ellv$, 
along with a function, $\veig(\ellv)$, as follows:
\beq
\calz_{GW}(\omega) = \int d\ellv \; \exp\left(
N^2 \left( \omega \ellv - \veig(\ellv) \right) \right) \; .
\label{legendre}
\eeq
$\ellv$ is just a complex number, so the 
stationary point of this integral occurs when
\beq
\omega = \left.
\frac{\partial \veig}{\partial \ell}\right|_{\ell = \ell_0} \; .
\label{stationary_leg}
\eeq

It is then trivial to see that if we choose
\beq
\veig(\ellv) = + \; \ellv^2 \;\;\; , \;\;\; \ellv \leq \frac{1}{2} \; ;
\label{VDM_less1/2}
\eeq
\beq
\veig(\ellv) = - \frac{1}{2} \log\left( 2\left( 1 - \ellv \right) \right)
+ \frac{1}{4} \;\;\; , \;\;\; \ellv \geq \frac{1}{2} \; .
\label{VDM_greater1/2}
\eeq
then (\ref{stationary_leg}) gives the
desired relations, (\ref{elln_omA}) and (\ref{elln_omB}).
One can also check that (\ref{legendre}) gives the
correct result for $\calz_{GW}(\omega)$ \cite{gross_witten}.

Using this form of $\veig$ in (\ref{legendre}), 
and replacing $\omega = i \overline{\omega}$,
the original partition function becomes
$$
\calz = \int d\lambda \int d\overline{\omega} \int d\ellv
\; \exp\left( - N^2 \, \calv' \; \right) \; , 
$$
\beq
\calv' = 
\vcl(\lambda) + i \overline{\omega}(\lambda - \ellv) + \veig(\ellv) \; .
\eeq
Doing the integral over $\overline{\omega}$ 
fixes $\lambda = \ellv$, and leaves
\beq
\calz = \int d\ellv \; \exp\left(- N^2\, \veff(\ellv) \right) \; .
\label{potential_eff}
\eeq
The effective potential,
\beq
\veff(\ellv) = \vcl(\ellv) + \veig(\ellv) \; ,
\label{eff_potential}
\eeq
is the sum of the loop potential, which we started with, and $\veig$.
Unlike the original integral over the matrix $\boldl$, 
(\ref{potential_eff}) is 
just an integral over a single degree of freedom, $\ellv$.
Since the effective potential is multiplied by an overall factor of $N^2$,
at large $N$
the true vacua of the theory are the stationary points of $\veff(\ellv)$.

When we introduce $\veig(\ellv)$ in (\ref{legendre}), 
mathematically this is just the Legendre transformation of
$\calz_{GW}$ \cite{zinn}, treating $\omega$ as an external source
for $\ellv$.  
This is nontrivial only because in $\calz_{GW}$,
the measure for the matrix $\boldl$ includes the Vandermonde determinant.
For this reason, we refer to $\veig(\ellv)$ as the ``Vandermonde potential''.  

Note that the Vandermonde potential is just the Legendre transformation
of $\calz_{GW}$, and not of the full integral.  One could compute
the Legendre transformation of $\cal Z$, but of necessity,
this would also include
the loop potential.  When the transition occurs, however, the
effective potential has degenerate vacua, and 
$\veff(\ell)$ is not a monotonic function of $\ell$.
This leads to well known ambiguities in 
the Legendre transformation \cite{zinn}.
In contrast, the Vandermonde potential is a monotonically
increasing function of
$\ell$, and there is never any ambiguity in its Legendre transformation.
This remains true at finite $N$, Sec.~\ref{three_colors}.

The Vandermonde potential
vanishes at $\ell = 0$, and diverges, logarithmically, as
$\ell \rightarrow 1$.  
At $\ellv = \frac{1}{2}$, the value of $\veig(\ell)$, and its
first and second derivatives, are continuous.  The
third derivative is not, since it vanishes for $\ell < \frac{1}{2}$,
and is nonzero for $\ell > \frac{1}{2}$.

For a lattice theory, $\omega$ is an inverse coupling constant.
With a Wilson action, there implies a transition, of third order,
in $\omega$.  This disappears if the lattice action is other
than Wilson, however.  
In contrast, the Vandermonde potential 
{\it always} has a discontinuity, of third order, at $\ellv = \frac{1}{2}$.
Because it is a transition of such high order, 
it may not affect the bulk thermodynamics significantly, but it is
always there.

The form of the Vandermonde potential is of interest.  
While it is nonpolynomial when $\ellv \geq \frac{1}{2}$,
$\sim \log(1 - \ellv)$, this is just eigenvalue repulsion
from the Vandermonde determinant.  
This is why there are no
constraints on the signs of the couplings in the loop potential:
eigenvalue repulsion never lets the value of $\ell$ exceed one.

About the origin, the Vandermonde potential is just a mass term,
but in fact that is the most interesting thing about it.
All terms of order $\sim \ellv^4$, $\sim \ellv^6$, and so on, conspire
to cancel \cite{kogut}.  In a miracle of group theory, the leading correction
is of very high order, starting out as $\sim \ellv^N$.  

Given the effective potential, it is then immediate to read off
the phase diagram of the theory.  For a given value of $m^2$
and the coupling constants, one just varies with respect to
$\ellv$, as one would for any other potential.   

For a given value of $m^2$, we denote the stable minimum as $\elln$.
The confined vacuum has $\elln = 0$:
if the confined vacuum is a stable minimum of
the loop potential, then because of the simple form of
the Vandermonde potential, it remains so for the effective
potential, with $\veff(0) = 0$.

A vacuum in the deconfined phase satisfies
\beq
\left. 
\frac{\partial \veff(\ell)}{\partial \ell}\right|_{\ell = \elln} = 0 \; ,
\eeq
with $\elln \neq 0$.  The transition occurs when 
the deconfined phase is degenerate with the confined phase, 
\beq
\veff(\ellnp) = 0 \; ;
\eeq
$\ellnp$ denotes the value of the order parameter at the transition,
approaching it in the deconfined phase.

The simplest possible example is to ignore {\it all} couplings
in the loop potential, and simply take
\beq
\vcl(\ellv) = - m^2 \; \ellv^2 \; .
\label{adjoint_potential}
\eeq

About the origin, the potential is
\beq
\veff(\ellv) = (- m^2 + 1) \; \ellv^2 \; \;\; , \;\;\; 
\ellv \leq \frac{1}{2} \; .
\label{gw_pot_effA}
\eeq
and
\beq
\veff(\ellv) = - m^2 \; \ellv^2 
- \frac{1}{2} \log\left( 2\left( 1 - \ellv \right) \right)
+ \frac{1}{4} \;\;\; , \;\;\; \ellv \geq \frac{1}{2} \; .
\label{gw_pot_effB}
\eeq
In the confined phase the potential is just a mass term.
Its sign shows that a transition occurs at $m^2 = 1$,
between a confined phase, for 
$m^2 < 1$, and a deconfined phase, for $m^2 > 1$.
In the deconfined phase, the condensate is the solution of
\beq
- 2 m^2 \elln + \frac{1}{2(1 - \elln)} = 0 \; ,
\label{soln_ellnp}
\eeq
which is
\beq
\elln = \frac{1}{2} \left( 1 + \sqrt{1 - \frac{1}{m^2}} \right) \; .
\label{solution_zero_field}
\eeq
(There is another solution to the quadratic equation, but it
has $\elln < \frac{1}{2}$, and so doesn't matter.)

The effective potential at the transition,
$m^2 = 1$, is shown in Fig.~\ref{fig:gw}.  We call this the
``Gross--Witten point''.  At the transition, 
the potential vanishes for $\ellv$ between $0$ and 
$\ellnp = \frac{1}{2}$; it then increases for $\ellv > \frac{1}{2}$.
Away from the Gross--Witten point, the potential is rather
ordinary.  In the confined phase, it is monotonically
increasing from $\ellv = 0$.  In the deconfined phase,
the global minimum has $\elln \neq 0$.
(For $m^2 > 1$, $\ellv = 0$ is actually metastable, since although
there is no barrier, the first derivative of $\veff$ vanishes at $\ell = 0$.)
As $m^2$ increases in value
from one, the minimum moves from $\frac{1}{2}$ to larger values,
approaching one only asymptotically in the limit of $m^2 \rightarrow \infty$.
For any value of $m^2$, either positive or negative,
there is only one stable minimum.

We can also compute the effective mass squared.  With this method,
it is simply the second derivative of the potential about the minimum,
\beq
m^2_{eff} = \left. 
\frac{\partial^2 \veff(\ell)}{\partial \ell^2}\right|_{\ellv=\elln} .
\label{define_eff_mass}
\eeq
In the confined phase, 
\beq
m^2_{eff} = 2 (1 - m^2) \;\;\; , \;\;\; m^2 \leq 1 \; .
\label{phys_mass_confined}
\eeq
In the deconfined phase, using (\ref{soln_ellnp}) we can write
\beq
m^2_{eff} = \frac{2 \elln - 1}{2 \elln (1 - \elln)^2} \;\;\; , \;\;\;
\elln \geq \frac{1}{2} \; .
\label{phys_mass_deconf}
\eeq
About the transition, 
\beq
m^2_{eff} \approx 4 \sqrt{m^2 - 1} + \ldots \;\;\; , \;\;\; 
m^2 \rightarrow 1^+ \; .
\label{phys_mass_deconf_limit}
\eeq
This difference occurs because at the Gross--Witten point,
the new minimum is right at the point where there is a discontinuity
of third order.  Both masses vanish at the transition, but do so
asymetrically, with different powers of $|m^2 - 1|$.

To go further, we need to make some assumption about the relationship
between the coefficients of the loop potential and the temperature.
We assume a mean field relation between the adjoint mass and the
temperature $m^2 -1 \sim T-T_d$, 
neglecting the variation of the coupling constants with temperature.
As the transition is approached in the confined phase,
$m^2_{eff} \sim T_d - T$ as $T \rightarrow T_d^-$,
while in the deconfined phase,
$m^2_{eff} \sim (T - T_d)^{1/2}$ for $T \rightarrow T_d^+$ \cite{dhlop}.

At the transition, the potential vanishes at both degenerate
minima, 
$\veff(0) = \veff(\ellnp) = 0$.  The derivative with respect to
$m^2$ is discontinuous, though, vanishing
in the confined phase, but nonzero in the deconfined phase,
\beq
\left. 
\frac{\partial \veff(\ell)}{\partial T}\right|_{T\rightarrow T_d^+}
\sim \left.
\frac{\partial \veff(\ell)}{\partial m^2}\right|_{m^2 \rightarrow 1^+}
= \frac{1}{4} \; ,
\label{latent_heat}
\eeq
which shows that the latent heat is nonzero, arising entirely
from the deconfined phase.  The transition is 
``critical'' first order \cite{dhlop}:
although masses vanish, the order parameter jumps.

Our discussion of the Vandermonde potential follows that of
Aharony {\it et al.} \cite{aharony}.  
They argue that if the loop potential contains 
operators such as $\tr \boldl^2/N$,
$\veff(\ell)$ can only be plotted for $\ell < \frac{1}{2}$.  
For potentials which are simply powers of the fundamental loop,
however, we assert that 
the effective potential is meaningful, and most illuminating
to plot, over its entire range, for $\ell: 0 \rightarrow 1$.
In Sec.~\ref{couplings_largeN} we outline 
how to compute the effective potential for arbitrary loop potentials.

\subsection{Nonzero Background Field} 
\label{BGfield}

Minimizing the effective potential is an elementary exercise in
algebra.  We begin with a background \zn field, taking
the loop potential to be
\beq
\vcl(\boldl) = - h \; \real \, \ellf - m^2 \, |\ellf|^2 \; .
\label{class_pot_nonzero_h}
\eeq

In this case, there is a nontrivial minimum of the
effective potential for any nonzero
value of $h$.  When $\elln \leq \frac{1}{2}$, it is 
\beq
\elln = \frac{h}{2 (1 - m^2)} \; ,
\label{backgd_h_less}
\eeq
while for $\elln \geq \frac{1}{2}$,
\beq
\elln = \frac{1}{2} \left(
1 - \frac{h}{2 m^2} + 
\sqrt{ \left(1 + \frac{h}{2 m^2}\right)^2 - \frac{1}{m^2} } \right) \; .
\label{backgd_h_greater}
\eeq

Consider the case where $h$ is infinitesimally small.
If $m^2 - 1$ is not $\sim h$, we can expand in $h$:
$\elln(h) \approx h/2$ when $\elln < \frac{1}{2} - O(h)$, while
\beq
\elln(h) \approx 
\elln(h=0)
+ \left( -1 + \frac{1}{\sqrt{1 - m^{-2} }} \right) \frac{h}{4 m^2} + \ldots
\eeq
for $\elln > \frac{1}{2} + O(h)$.  
Thus away from $\elln \sim \frac{1}{2}$, on both sides of the transition
the shift in $\elln$ is small, of order $\sim h$ (see e.g.\ Fig.~2
in~\cite{loop2e}).
When $m^2 - 1 \sim h$, though, the solution differs by a large
amount from that for $h=0$.  There is a third order
transition when $\ell_0$ equals $\frac{1}{2}$.
The value at which this happens 
is easiest to read off from (\ref{backgd_h_less}).  This occurs for
$m^2 = 1 - h$, at which point the effective mass squared 
$m^2_{eff} = 2 h $.  

For an ordinary first order transition, as the value of 
a background field increases
from zero, there is still a nonzero jump in the order parameter.
The jump only disappears for some
nonzero value of the background field, which is a critical endpoint.
For larger values, there is no jump in the order parameter, and
masses are always nonzero.
For the loop potential of (\ref{class_pot_nonzero_h}),
however, any nonzero background field,
no matter how small, washes out the first order
transition.  For any $h \neq 0$, there is still a transition of
third order, at which the masses are nonzero.

These results give an elementary interpretation of the calculations
of Schnitzer \cite{schnitzer}.
He computed the partition function for quarks in the fundamental
representation coupled to $SU(\infty)$ gauge fields, taking
space to be a very small sphere.  Fields in the fundamental
representation break the \zn symmetry, and act like a background
field $h \sim N_f/N$, where $N_f$ is the number of flavors.
When $m^2 < 1 - h$, by (\ref{backgd_h_less})
$\elln \sim h \sim N_f/N$; substituting this back into
(\ref{class_pot_nonzero_h}), the potential, and so the free energy, is of 
order $\veff(\elln) \sim h^2 \sim (N_f/N)^2$.
The first order transition disappears for any $h \neq 0$ \cite{green_karsch},
leaving just a transition,
of third order, when $\ell_0$ passes through $\frac{1}{2}$ 
\cite{damgaard_patkos}.

\subsection{Mean Field Approximation}
\label{sec_mean_field}

The analysis in Secs.~\ref{gross_witten_point}
and~\ref{BGfield} is most transparent for understanding
the physics.  Following 
Kogut, Snow, and Stone \cite{dhlop,kogut,green_karsch}, we 
show how to compute the partition function in another way.
This was originally derived as a mean field approximation,
although for loop models, we show that 
it is equivalent to a large $N$ approximation.
We consider the case where the loop potential includes
just an adjoint loop; we also add a background field,
taking the loop potential to be that of (\ref{class_pot_nonzero_h}). 
The case of more general potentials follows directly, and is
addressed in Sec. IV for $N=3$.

In the expression of (\ref{constrained_pot}), 
$\lambda$ only appears in the loop potential and in
the constraint.  Thus before doing the matrix integral, we
can extremize with respect to $\lambda$.  
With the $\vcl$ of (\ref{class_pot_nonzero_h}), this fixes
$\lambda = (\omega - h)/(2 m^2)$.
We then define a potential directly from the matrix integral:
\beq
\exp(- N^2 \widetilde{\calv}(\omega) ) 
= \int d\boldl \exp(N^2 \omega \; \real \, \ellf) \; .
\label{single_site_pot}
\eeq
This leaves one remaining integral, over $\omega$.   
In fact, since the relationship between $\lambda$ and $\omega$
is linear, we can substitute one for the other, and so obtain
the mean field potential, 
\beq
\vmf(\ellv) = m^2 \ellv^2      
+ \widetilde{\calv}(2 m^2 \ellv + h) \; .
\label{mean_field_pot}
\eeq
To follow previous notation, we have also relabeled $\lambda$ as $\ellv$.

To understand the relationship to a mean field approximation, 
consider a lattice theory in which there are fundamental loops on each site,
coupled to nearest neighbors with strength $\sim m^2$.
The number of nearest neighbors also enters, but 
for our purposes, scales out.
Start with a given site, and assume that on adjacent sites,
all matrices have some average value, 
$\ell = \tr \boldl/N$.  The action is then the 
coupling, $m^2$, times the average value $\ell$,
times the value for that site, 
or $\sim 2 m^2 \ell \, \real \, \tr \boldl/N$ altogether.
In a background field, one also adds 
$\sim h \real \, \tr \boldl/N$.  
With $\omega = 2 m^2 \ell + h$, this is
the integral of (\ref{single_site_pot}), and 
gives $\widetilde{\cal V}$.  The remaining part of the 
mean field potential, $m^2 \ellv^2$, 
arises by imposing the mean field condition that the
expectation value of $\real \, \boldl$ equal the presumed average value,
$\ell$ \cite{kogut}.

Several aspects of the mean field approximation are much
clearer when viewed as the large $N$ expansion of a loop potential.
We see that the mean field action, involving the
real part of the fundamental loop, actually arises from a loop
potential with an adjoint loop, 
$= m^2 |\ellf|^2$.  Thus if we add an additional 
term proportional to an adjoint loop, 
$\widetilde{m}^2 |\ellf|^2$, to the mean field action, 
$ = 2 m^2 \ell \, \real \, \ellf$, this is equivalent to
a shift in the coupling for the adjoint loop,
$m^2 \rightarrow m^2 + \widetilde{m}^2$.  
This was proven previously in \cite{dhlop}, by more indirect means.

For zero background field, the mean field potential is
\beq
\vmf(\ellv)
= m^2 (1 - m^2) \ellv^2 \;\;\; , \;\;\; \ellv \leq \frac{1}{2 m^2} \; ,
\label{pot_mf_less}
\eeq
and
\beq
\vmf(\ellv)
= - 2 m^2 \ellv + m^2 \ellv^2 + 
\frac{1}{2} \log (2 m^2 \ellv) + \frac{3}{4} 
\; , \; \ellv \geq \frac{1}{2 m^2} \; .
\label{pot_mf_greater}
\eeq

Notice that the mean field potential, $\vmf(\ellv)$, 
does not agree with the previous form of the effective potential,
$\veff(\ellv)$ in
(\ref{gw_pot_effA}) and (\ref{gw_pot_effB}).  Even the
point at which 
the mean field potential is nonanalytic, $\ell = 1/(2m^2)$, is not
the same.

Although the mean field and effective potentials differ, they
do give the same vacua.
In the deconfined phase, the stationary point of (\ref{pot_mf_greater}) 
satisfies
\beq
-2 m^2 (1 - \elln) + \frac{1}{2 \elln} = 0 \; ,
\eeq
which agrees with (\ref{soln_ellnp}).  One can also
check that for $h \neq 0$, the solution coincides with 
(\ref{backgd_h_less}) and (\ref{backgd_h_greater}).

Second derivatives of the mean field
and effective potentials are not the same,
even at the correct vacuum $\elln$.
For instance, in the confined phase the second derivative of
(\ref{pot_mf_less}) is $= 2 m^2 (1-m^2)$, not $= 2(1-m^2)$
of (\ref{phys_mass_confined}).  They agree 
to leading order in $1-m^2$
about $m^2 \rightarrow 1$, but not otherwise.
Likewise, in the deconfined phase, the second derivative of
$\vmf$ only agrees with that of $\veff$, (\ref{phys_mass_deconf_limit}),
to leading order in $\sqrt{m^2 -1}$ as $m^2 \rightarrow 1$.

To understand this discrepancy, we define a mass not by
the potential {\it per se}, but by the response to a background
field.  Computing the partition function in the presence of $h\neq 0$,
the second derivative is
\beq
\left. \frac{1}{N^4} \frac{\partial^2}{\partial h^2} \log \,\calz(h)
\right|_{h=0} = 
\langle (\real \; \ellf)^2 \rangle -
\langle \real \; \ellf \rangle^2 \; .
\label{part_back_h}
\eeq
The great advantage of the approach of 
Sec. \ref{gross_witten_point} is that the background field 
$h$ only appears linearly in the action, through the loop potential.
In this case, (\ref{part_back_h}) 
$= -1/(2 N^2 m^2_{eff})$, where $m^2_{eff}$ is just the
second derivative of the effective potential, (\ref{define_eff_mass}).

In the mean field approach, however, $h$ enters into
$\widetilde{\cal{V}}(2 m^2 \ell + h)$; this function
has terms of both linear {\it and} quadratic order in $h$.
In order to compute (\ref{part_back_h}), then, it is
necessary to include the terms quadratic in $h$.
This can be done, but is tedious.  The result 
is that the effective mass, defined properly from 
(\ref{part_back_h}) and computed with $\vmf$, agrees with that  
obtained so easily from $\veff$, 
(\ref{phys_mass_confined}) and (\ref{phys_mass_deconf}).  

In the presence of a nonzero
background field, following Damgaard and Patkos \cite{damgaard_patkos}
we have checked that 
there is a transition of third order when
$\elln$ passes through $\frac{1}{2}$.  
Even this is not obvious for the mean field potential,
since the point at which $\vmf$ is nonanalytic 
depends upon the value of $m^2$.

In the end, the two methods must agree for physical quantities.
One is only doing the integrations in a different
order, and there are no subtleties in the integrands.
On the other hand, in the next
section we see that at finite $N$, the mean field approach is
advantageous for both formal calculations in the confined phase, 
and for numerical computation.

\subsection{Away from the Gross--Witten Point}
\label{away_gross_witten}

We now return to the case of zero magnetic field, and consider
higher interactions.  Consider a quartic potential \cite{aharony},
\beq \label{four_point_potential}
\vcl(\boldl) = - m^2 \, |\ellf|^2 + \kappa_4 \left( |\ellf|^2 \right)^2
\eeq
It is immediate to compute the phase diagram.
There is a 
line of second order transitions for $\kappa_4 > 0$, along
$- m^2 + 1 = 0$, and a line of first order transitions
for $\kappa_4 < 0$, along some line where
$- m^2 + 1 > 0$.  These meet
at what appears to be a tri-critical point,
where $\kappa_4 = -m^2 + 1 = 0$: this is the Gross--Witten point.

It is not a typical tri-critical point, however.  
If this were an ordinary scalar field, then 
$\ellnp$, the value of the order parameter
just above the transition, 
would be zero along the second order line,
zero at the tri-critical point, 
and then increase from zero as one moves away from the
tri-critical point along the first order line.

For the matrix model, however, $\ellnp = 0$
along the second order
line, but then jumps --- {\it discontinuously} --- to
$\ellnp = \frac{1}{2}$ at the Gross--Witten point.  

For example, for 
small and negative values of $\kappa_4$, 
the transition occurs when 
\beq
m^2 \approx 1 - \frac{|\kappa_4|}{4} + \ldots \; .
\label{msq_small_kappa}
\eeq
At the transition, $\ellnp$ increases from $\frac{1}{2}$ as
\beq
\ellnp \approx \frac{1}{2} + \frac{\sqrt{|\kappa_4|}}{4} + \ldots \; .
\label{vev_small_kappa}
\eeq
The masses are always nonzero; at the transition, in the confined phase,
\beq
m^2_{eff} \approx \frac{|\kappa_4|}{2} + \ldots \; ,
\eeq
while in the deconfined phase,
\beq
m^2_{eff} \approx \sqrt{|\kappa_4|} + \ldots \; .
\eeq
When $\kappa_4 \rightarrow 0$, the mass in the confined phase
vanishes more quickly than in the deconfined phase.  This is similar to
what happens when $\kappa_4 = 0$, 
(\ref{phys_mass_confined}) and (\ref{phys_mass_deconf_limit}).

Our analysis differs from that of
Aharony {\it et al.} \cite{aharony}.  The caption of their Fig. 3
is correct, stating that for small $\kappa_4$,
the transition occurs at (\ref{msq_small_kappa}).
In the Figure, however, the curve for this $m^2$ 
is actually in the deconfined phase.
This is not evident, since they only plot
the potential for $\ell < \frac{1}{2}$.  It can be seen
by noting that at the transition, $\veff(\ellnp)=0$; as
$\ellnp > \frac{1}{2}$, (\ref{vev_small_kappa}), then
$\veff(\frac{1}{2}) > 0$, and not zero, as in their Figure.

Moving to increasingly negative values of $\kappa_4$, the
value of $\ellnp$ increases as well, as do the masses.
For example, consider $\kappa_4 = - 1$.
Numerically, we find that 
the transition occurs when $m^2 \approx 0.456121...$, with 
$\ellnp \approx .841176...$.  
The effective potential
at the transition is illustrated in
Fig.~\ref{fig:first}.  The masses are
very different in the two phases: in the confined
phase, $m^2_{eff} = 1.08776...$,  while in the deconfined phase, 
$m^2_{eff} = 10.4184...$.  Thus, for such a large value of
$- \kappa_4$, the transition appears to be a perfectly ordinary,
strongly first order transition.  The masses are always nonzero,
including at the transition point.  In Fig.~\ref{fig:first},
the potential does have a discontinuity, of 
third order, at $\ell = \frac{1}{2}$; however, 
as $\ellnp > \frac{1}{2}$, this is of little consequence.

We remark that because the masses are very different in the two
phases, the potential in Fig.~\ref{fig:first} cannot be approximated
by a quartic potential in $\ell$.  We return to this for
$N=3$ in Sec. \ref{three_colors}.

As $\kappa_4 \rightarrow - \infty$, $m^2 \approx
|\kappa_4|$, and $\ellnp \approx 1 - 1/4|\kappa_4|$.
Due to eigenvalue repulsion in the Vandermonde potential,
the expectation value is always less than unity.
Since $\ellnp \rightarrow 1$ as $\kappa_4 \rightarrow - \infty$, 
in this case the deconfined phase is arbitrarily close to a truly
perturbative gluon plasma from temperatures of $T_d^+$ on up,
with the transition as strongly first order as possible.
It is reasonable that what is a strong coupling phase in the loop
model corresponds to a weakly coupled regime in the underlying
gauge theory.  

Next, consider adding six-point interactions, 
$\sim \kappa_6 (|\ellf|^2)^3$.  For the purposes of
discussion, we take $\kappa_6$ to be positive, although
this is not necessary.  For $\kappa_6 > 0$, the phase
diagram, in the plane of $m^2$ and $\kappa_4$, is greatly
altered.  There is a second order line when $\kappa_4 > 0$,
which meets a first order line for $\kappa_4 < 0$, but they
meet at a true tri-critical point, for $\kappa_4 = 0$.  
At this tri-critical point, the jump in the order parameter vanishes,
as does the mass.  

With only quartic interactions, either $\ellnp = 0$
or $\ellnp \geq \frac{1}{2}$ \cite{aharony}.  This 
is no longer true when six-point interactions are included.
Consider the loop potential
\beq
\vcl(\boldl) = \kappa_6 |\ellf|^2 \left( |\ellf|^2 - \ell_c^2 \right)^2
+ (m^2 - 1) |\ellf|^2 \; ,
\label{six_point}
\eeq
where $\ell_c$ is some number.
If $\ell_c < \frac{1}{2}$, then there is a first order
transition when 
$m^2 = 0$, as $\ellnp = \ell_c$ at the transition.
If $\ell_c > \frac{1}{2}$, then one must use the Vandermonde
potential in (\ref{VDM_greater1/2}), and while the transition
is still of first order, it does not occur when $m^2 = 0$.
We do not work out specific examples, since we only wanted
to make the point that it is possible to have first order transitions
in which $\ellnp$ assumes any value between zero and one. 

In the cases in which there is a second order transition,
such as $\kappa_4 > 0$, there is nothing remarkable about it:
it is in the universality class of a $U(1)$ spin, and exhibits
standard critical behavior.

In all cases, if $\ellnp < \frac{1}{2}$, then there is a third
order transition when the value of $\elln$ passes through
$\frac{1}{2}$ \cite{aharony}.  
Since the masses are nonzero at the transition, this third order
transition does not appear to be of much consequence.
If $\ellnp > \frac{1}{2}$, there is no third order transition,
just one first order transition.

The only way to have a first order transition, in which
both masses vanish,
is if {\it all} couplings vanish: $\kappa_4 = \kappa_6 = \ldots = 0$.
For this reason, 
we term the Gross--Witten point an ultra-critical
point, since an infinite number of couplings must be tuned to
zero in order to reach it.  

\subsection{Couplings at Large N}
\label{couplings_largeN}

In this subsection we 
comment on how couplings in the loop potential scale with $N$.

To one loop order in the perturbative regime, terms such as
as terms $\sim \tr \boldl^2/N$ arise in the loop potential
\cite{previous,aharony,fss,schnitzer,aharony2,interface1,interface2}.
This is $\sim 1$ at large $N$, and so 
allowed.  In terms of normalized loops, however, 
$\tr \boldl^2/N \sim N(\ell_{N^2} - \ell_N^2)$,
where $\ell_{N^2}$ is a loop for a two-index 
tensor representation, Eq. (42) of \cite{dhlop}.
Thus $|\tr \boldl^2/N|^2 \sim N^2|\ell_{N^2} - \ell_N^2|^2$.
Traces such as 
$|\tr \boldl^p/N|^2$, which are also allowed in the loop potential,
are $\sim N^{2(p-1)}$ times differences of normalized loops.

For the general loop potential of (\ref{general_potential}),
the Gross--Witten point occurs when
$m^2 =1$, with all other couplings $\kappa_\cals=0$.
Thus we studied a weak coupling regime close
to the Gross--Witten point, where the $\kappa_\cals \sim 1$
at large $N$.  A regime in which the couplings 
$\kappa_\cals$ grow
with powers of $N$ is one of strong coupling,
far from the Gross--Witten point.

For strong coupling, the loop 
potential depends not just upon the fundamental loop,
but on all 
$N-1$ degrees of freedom of the matrix $\boldl$.
A convenient choice would be the fundamental loop, $\tr \boldl/N$, 
plus $\tr \boldl^2/N$, $\tr \boldl^3/N$,
$\ldots \tr \boldl^{N-1}/N$.  It is then necessary
to compute all potentials in the associated $N-1$ dimensional space.  
This will be involved, since an expectation value for the fundamental
loop automatically induces one for $\tr \boldl^2/N$, {\it etc.}

Aharony {\it et al.} studied this strong coupling regime by adding
$|\tr \boldl^2/N|^2$ to the loop potential \cite{aharony}.  
Using the solution of Jurkiewicz and Zalewski \cite{jurk}, 
they find that 
the behavior is similar to that in weak coupling when
$\kappa_4 \neq 0$.  The bulk thermodynamics only
exhibits ordinary first or second order transitions, plus
third order transitions when $\elln$ passes through $\frac{1}{2}$.
This is most likely valid everywhere in the strong coupling regime.

Thus whether or not deconfinement is near the Gross--Witten point
when $N \geq 3$, in the space of couplings $\kappa_\cals$,
it is manifestly the most interesting place it could be.

\section{Finite N} 
\label{sec_finiteN}

\subsection{Z(N) Neutral Loops in the Confined Phase}
\label{neutral_loops}

What made our treatment at infinite $N$ so simple is the
assumption that because of factorization, the only (normalized)
loops which enter are those for the fundamental representation.
At finite $N$, one cannot avoid considering loops in higher representations.

We consider only effects of the loop potential, neglecting kinetic
terms.  While this is exact at infinite $N$, it is not justified
at finite $N$.  Thus the following analysis should be considered
as the first step to a complete, renormalized theory, including
the kinetic terms of Sec.~\ref{matrix_models_loops}.

In this subsection we begin with some general observations about
the expectation values of loops in arbitrary representations:
\beq
\langle \ell_\calr \rangle = \int d\boldl \;\;
\left. \ell_\calr \; \exp\left(- N^2 \vcl(\boldl)\right)\right/\calz
\; ,
\label{arb_exp_val}
\eeq
with $\calz$ that of (\ref{partition_function}).

We concentrate on the stationary point of the partition function.
This classical approximation is exact at infinite $N$;
how well it applies at finite $N$, even to constant fields, is
not evident.  For $N=3$, however, the overall factor of $N^2$ 
[or in fact, $N^2 -1 = 8$, cf.\ eq.~(\ref{partition_function_B})] in the
exponential suggests that this might be reasonable.  While
less obvious for $N=2$, we suggest later a very specific test
which can be done through numerical simulations.

In fact, we are really forced to consider just the stationary point
of the partition function.  
If we did the complete integral, then at finite $N$ the \zn symmetry
would never break spontaneously.  
In a phase which we
thought was deconfined, all \zn transforms of a given vacuum would
contribute with equal weight, so that in the end, all \zn charged
loops would vanish (\zn neutral loops would be nonzero).
This is just the well known fact that a symmetry
only breaks in an infinite volume, or at infinite $N$ \cite{aharony}.
By taking the stationary point of the integral, we are forcing it
to choose one of the vacua in which the \zn symmetry is broken.

As noted before, the matrix $\boldl$
depends upon $N-1$ independent eigenvalues.  
Instead of the eigenvalues, we can choose the $N-1$ independent quantities 
to be $\tr \boldl$, $\tr \boldl^2$, and so on, up to $\tr \boldl^{N-1}$.  

We prefer to deal with a set of loops.  The first trace is of course the
fundamental loop, $\ellf = \tr \boldl/N$.
Any potential is real, and so we also choose the anti-fundamental 
loop, $\ell_{\overline{N}} = \ellf^*$.  
As a Young tableaux, the anti-fundamental
representation corresponds 
to the anti-symmetric tensor representation with $N-1$ fundamental
indices.

There is no unique choice for the remaining loops, but as
the anti-fundamental representation is a purely anti-symmetric
representation, we can chose to work with only anti-symmetric
representations.  Thus 
we choose loops for anti-symmetric tensor representations with
$j$ indices, $j = 1$ to $N-1$, denoting this set as $\calra$.

For $N=2$, $\calra$ is just the doublet loop, which is real.  
For $N=3$, it is the triplet and anti-triplet loops.
For $N=4$, $\calra$ includes the quartet, anti-quartet,
and sextet loops.  The quartet 
transforms under a global symmetry of $Z(4)$, 
while the sextet only transforms under $Z(2)$; both are needed,
to represent the case in which only $Z(2)$ breaks, but $Z(4)$ doesn't.
Including all of the loops in $\calra$ ensures that all possible
patterns of \zn symmetry breaking can be represented for $N \geq 4$.

The previous form of the loop potential,
(\ref{general_potential}), is an 
infinite sum of loops in 
\zn neutral representations, the $\calrn$.
Instead, we now 
take the potential to be a function only of the loops in $\calra$, 
\beq
\vcl(\boldl) = \vcl(\ell_\cals) \;\;\; , \;\;\;
\cals \; \epsilon \; \calra \; .
\label{potential_sum_calra}
\eeq
Previously, the potential was a sum over the loops in $\calrn$, each
appearing only to linear order.  Now, only the $N-1$ loops in
$\calra$ enter, but these do so through all \zn invariant polynomials,
to arbitrarily high order.

This is useful in introducing constraints into the partition function.
One might have thought that it is necessary to introduce constraints
for all fields which condense, but in fact, it is only necessary
to introduce constraints for fields which appear in the action \cite{zinn}.
Even so, if the action is written in terms of the $\calrn$, there are
an infinite number of such loops.  The advantage of the $\calra$
is that we only have to introduce $N-1$ constraints.

As usual, we start by introducing constraints into the partition function, 
\beq
\calz = \int d\boldl \; \prod_{\cals \epsilon \calra}
\int d\lambda_\cals \;\;
\delta(\lambda_\cals - \ell_\cals) \; 
\exp\left( - N^2 \vcl(\lambda_\cals) \right) \; .
\eeq
where in the potential we have used the delta-functions to
replace $\vcl(\ell_\cals)$ by $\vcl(\lambda_\cals)$.
Introducing constraint fields,
\beq
\calz = \int d\boldl \; \prod_{\cals \epsilon \calra}
\int d\lambda_\cals \; \int d\omega_\cals\;
\exp\left( - N^2 \calvc \right) \; ,
\eeq
we obtain the constraint potential,
\beq
\calvc = \vcl(\lambda_\cals) + \sum_{\cals \epsilon \calra}
i \, \omega_\cals \left( \lambda_\cals - \ell_\cals \right) \; .
\eeq
(In this subsection alone we write $\omega_\cals$ instead of
$\overline{\omega}_\cals$, since all $\omega_\cals$'s vanish
at the confined stationary point.)
We now follow the mean field approach, defining the potential
\beq
\exp(- N^2 \widetilde{\calv}(\omega_\cals) ) 
= \int d\boldl \; \exp \left( N^2 \sum_{\cals \epsilon \calra}
\omega_\cals \; \ell_\cals \right) \; .
\label{general_single_site}
\eeq
After integrating over $\boldl$, we obtain
\beq
\calz = \prod_{\cals \epsilon \calra}
\int d\lambda_\cals \; \int d\omega_\cals\;
\exp\left( - N^2 \vmf(\lambda_\cals,\omega_\cals) \right) \; ,
\label{general_part_fnc}
\eeq
where
\beq
\vmf(\lambda_\cals,\omega_\cals) = 
\vcl(\lambda_\cals) + \sum_{\cals \epsilon \calra}
i \, \omega_\cals \lambda_\cals
+ \widetilde{\calv}(i \omega_\cals) \; .
\eeq
This is the mean field potential, written in terms of the $\calra$ loops.

By an overall \zn rotation, we can always assume
that the stationary point is real.  Thus in order
to determine just the stationary point, only the real part of
any loop will enter.  The integral which determines $\widetilde{{\cal V}}$
for the fundamental loop is familiar : see Table
12 of \cite{drouffe_zuber} and \cite{epsilon}.  The integral 
in (\ref{general_single_site}) is more general, involving
all loops in $\calra$. 

We use this formalism to prove an elementary theorem which is mathematically
trivial, but physically important.  The obvious guess for
the confined phase is where the expectation values of all loops
in the action vanish, 
$\lambda_\cals = 0$ for all $\cals$ in $\calra$.  
We can always define the loop potential so that it vanishes
when all $\ell_\cals$ vanish; any constant term in $\vcl$ drops
out of the ratio in (\ref{arb_exp_val}).
Further, as the potential is \zn neutral, and as all $\lambda_\cals$
carry \zn charge, the first derivative of $\vcl$ with
respect to any $\lambda_\cals$ has \zn charge,
and vanishes if $\lambda_\cals = 0$.
Thus all constraint fields vanish at the
stationary point, $\omega_\cals = 0$.  
\zn invariance also implies that 
the first derivative of $\widetilde{{\cal V}}(\omega_\cals)$
with respect to any $\omega_\cals$ vanishes.
Altogether, $\vmf = 0$ in the confined phase,
$\lambda_\cals = \omega_\cals = 0$, and this point is extremal.
If the potential vanishes, though, the expectation value of any loop
is simply an integral over the invariant group measure.  For
any (nontrivial) representation, however, 
whatever the \zn charge of the loop, its integral 
over the invariant measure vanishes identically,
$ \int \; \ell_\calr \; d\boldl = 0$.
Consequently, the expectation values of {\it all} loops vanish:
\beq
\langle \ell_\calr \rangle = 0 \;\;\; , \;\;\; T \leq T_d^-
\label{loops_vanish}
\eeq
While the confined vacuum is extremal, stability is
determined by second derivatives of the potential, and only holds
for $T \leq T_d^-$.

This is very unlike what would be expected merely on the basis
of \zn symmetry.  For \zn neutral loops, such as the adjoint, 
there is no symmetry which prohibits their acquiring a
nonzero expectation value in the confined phase.  We stress
that this result is valid only in mean field theory, 
when fluctuations from kinetic terms are neglected.

For three colors, lattice data indicates that the expectation
value of the renormalized
adjoint loop is very small below $T_d$ \cite{dhlop}.
Since the deconfining transition is of first order for three
colors (in four spacetime dimensions), this is not the best place
to look for the expectation value of \zn neutral loops below $T_d$.
Rather, it is preferable to study a deconfining
transition of second order, 
as occurs for $N=2$ in three or four
spacetime dimensions, and even for $N=3$ in three spacetime dimensions.
We remark that for a second order transition, universality
implies that two point functions, such as
$\langle \ellad(x) |\ellf(0)|^2 \rangle$, scale with the
appropriate anomalous dimensions.  Universality, however,
places no restriction on the one point functions of \zn neutral loops.
Merely on the basis of \zn symmetry, one does not expect them to
be small in the confined phase: this is a clear signal that
the underlying dynamics is controlled by a matrix model, and not
just by some type of effective \zn spin system.

In this regard, Christensen and Damgaard,
and Damgaard and Hasenbusch \cite{damgaard}, 
also noted that in a classical approximation, 
not only do the expectation values
of \zn neutral loops vanish in the confined phase, but for
a second order 
deconfined phase, they vanish like a power of the fundamental loop.
As $T\rightarrow T_d^+$,
$\langle \ell_\calr \rangle \sim 
\langle\ellf\rangle^{p_+} \langle\ellf^*\rangle^{p_-}$, up
to corrections involving higher powers of the fundamental and
anti-fundamental loops.  The powers are identical to those
of factorization at large $N$ \cite{dhlop}, but
the coefficient is not the same; to satisfy factorization,
the coefficient is one, up to corrections of order $1/N$.  
For $N=2$, it is $2/3$.

We conclude this section with a comment.
There is always a given element of $SU(N)$ for
which the trace of the fundamental loop vanishes; {\it e.g.}, in
$SU(2)$ it is the diagonal matrix $\pm (1,-1)$.
Thus one might imagine modeling the confined vacuum as an
expansion about such a fixed element of the group group \cite{ogilvie}.
Because these matrices are not elements
of the center of the group, however, they are 
not invariant under local $SU(N)$ rotations: they represent
not a confined, but a Higgs, phase.  This is very different
from the above, where in the confined phase,
one integrates over all elements of the gauge group with
equal weight.  While the expectation
value of the fundamental loop vanishes in a Higgs phase, 
those of higher representations
do not; that of the adjoint loop $=- 1/(N^2 - 1)$.
This does not agree with lattice simulations for $N=3$, 
which find a very small value for the octet loop below $T_d$,
$\ll 1/8$ \cite{dhlop}.

\subsection{Three Colors}
\label{three_colors}

For $N=3$, the analytic solution of large $N$ is not available.
There are various approximation schemes which one can try.

For example, one can expand about the confined phase, $\ell = 0$.
{}From Eq.~(2.22) of \cite{kogut}, however, one can see
that while such an expansion works well for small $\ell \leq 0.2$,
it fails at larger values, and so is not of use near
the Gross--Witten point.  
(Note that \cite{kogut} absorbs the number of nearest neighbors
into their parameter $\alpha$, so $\alpha = 6\ell$;
also, their $\beta = (N^2-1) m^2/N^2 = 8m^2/9$ due
to our definition of the prefactor of $\vcl$ in
eq.~(\ref{partition_function_B}).)  

One can also expand about the deconfined phase,
taking $m^2 \rightarrow \infty$, which forces $\ell \rightarrow 1$.
Since for $N=3$ we need only consider the triplet loop, 
the integral 
for the mean field potential, (\ref{general_single_site}), is 
the $N=3$ analogy of (\ref{single_site_pot}).  At the stationary
point, $\omega = 2 m^2 \ell$, so at large $m^2$, where
$\ell \approx 1$, $\omega$ is also large.
Mathematically, the integral for the mean field potential is
the same as arises in strong coupling expansions 
of a lattice gauge theory.  The result for large $\omega$ 
is given in Table 12 of \cite{drouffe_zuber}.  
The result starts as $\sim \omega$,
plus a term $\sim \log(\omega)$, and then
a power series in $1/\omega$.  The first term is
trivial, due to the fact that $\ell \approx 1$.
The term $\sim \log(\omega)$ 
represents eigenvalue repulsion from the Vandermonde
determinant.  This was seen before
at large $N$, as the term $\sim \log(1-\ell)$ in the Vandermonde
potential, (\ref{VDM_greater1/2}).  
One finds that while this expansion works well
near $\ell \approx 1$, it does not appear useful for
smaller values.  

The failure of these perturbative expansions is not particularly
remarkable.  In terms of $\ell$, the Gross--Witten point is
identically midway between a confined, and a completely deconfined,
state.  There is no reason why an expansion about either limit should
work, although they might have.  Even so, 
for $N=3$ the matrix model is just a two dimensional
integral.  The regions of integration are finite,
and there are no singularities in any integral.  Thus 
we can just do the integrals numerically.  

We define the partition function as
\beq
\calz = \int d\boldl \; \exp\left( - (N^2 - 1) \vcl(\boldl) \right) \; .
\label{partition_function_B}
\eeq
We multiply the loop potential by $N^2 - 1$, instead of $N^2$.
For the loop potential, this is a matter of convention, but
for the Vandermonde potential, numerically we find that with
this definition, the $N=3$ results are closer to those of 
$N=\infty$.
This is not surprising: in expanding perturbatively about the deconfined
state with $\ell = 1$, an overall factor of the number of generators,
which for an $SU(N)$ group is $N^2 - 1$, 
arises naturally \cite{drouffe_zuber}.

By the previous section, we consider the loop potential as
a function of the triplet and anti-triplet loops.  The most general
potential is then
\beq
\vcl(\boldl) =
- m^2 |\ellt|^2 + \kappa_3 \; \real \; (\ellt)^3 
+ \kappa_4 \left( |\ell_3|^2 \right)^2 + \ldots 
\label{three_gen_pot}
\eeq
These terms represent octet, decuplet, and $27$-plet representations
of $SU(3)$.  Other terms of higher order in $\ellt$ include
one of pentic order, and two of hexatic order \cite{loop2c}.

We then mimic the analysis of Sec. \ref{gross_witten_point} to obtain the
Vandermonde potential, $\veig(\ell)$.  The only difference is
that at large $N$, the expectation value of $\ellf$, in the presence of an
external source, is known analytically,
(\ref{elln_omA}) and (\ref{elln_omB}).  For $N=3$, it is necessary to
determine this relationship numerically, from the integral
\beq
\ell(\omega) = \int d\boldl \;
\real \; \ellf \; \exp( (N^2 - 1) \; \omega \; \real \, \ellf) \; ,
\eeq
where $\omega$ is a real source.
Given $\ell(\omega)$, it is then trivial to
invert this function, to obtain the source as a function 
of the expectation value, $\omega(\ell)$.  
Following (\ref{stationary_leg}),
the Vandermonde potential is then just the integral of the
source, with respect to $\ell$:
\beq
\veig(\ell) = \int^\ell_0 d\ell' \; 
\omega(\ell')\; ,
\eeq

As at infinite $N$, when $N=3$ the Vandermonde potential
is a monotonically increasing function of $\ell$: it vanishes at $\ell = 0$, 
and diverges, logarithmically, as $\ell \rightarrow 1$.
Because this function is monotonically increasing, if we
know $\ell(\omega)$, then there is no ambiguity
in inverting it, to obtain $\omega(\ell)$.  

The numerical result for $N=3$ is compared in Fig.~\ref{fig:Vdm} to
the analytical expression for $N=\infty$ from 
eqs.~(\ref{VDM_less1/2}) and (\ref{VDM_greater1/2}).
The $N=3$ result is always less than that for $N=\infty$,
but even up to $\ell \approx 0.8$, they lie within a few percent of
one another.  This is most surprising: the $U(1)$ symmetry of
$N=\infty$ is broken at finite $N$ to \zn by operators
which start as $\ell^N$.  
For $N=3$, this is an operator of low dimension, $\sim \ell^3$.   
Indeed, in Fig.~\ref{fig:Vdm} we also show the result for
the $N=2$ Vandermonde potential, and find that even that isn't
so far from the $N=\infty$ result.

To obtain the effective potential, we simply add the loop potential
$\vcl$ to $\veig$. The result for the three-color analogue of the
Gross-Witten point, i.e.\ the potential from eq.~(\ref{three_gen_pot})
with all couplings $\kappa_i=0$, is depicted in Fig.~\ref{fig:VeffN=3}
by the solid line. The value of the order parameter at the transition,
$\ellnp \approx 0.485$, is very close to the $N=\infty$ value of
$\frac{1}{2}$ \cite{dhlop,kogut}.  Since the Vandermonde potential
for $N=3$ is so close to $N=\infty$, this small shift, by $\approx 3\%$,
is reasonable.  Similarly, 
as can be seen graphically, the potential is almost flat at $T_d$.
Masses are small at the transition, and there is a small
barrier between degenerate minima of the potential.
Physically, this corresponds to a nearly vanishing 
interface tension between phases. 

As mentioned in the Introduction, lattice data for $SU(3)$
Yang-Mills indicates that $\ell_0(T_d^+)$ is less than $\frac{1}{2}$, perhaps
$\approx 0.4\pm10\%$~\cite{dhlop}. Additional lattice simulations are
needed to fix the jump in the triplet loop 
more precisely. Nevertheless, in what follows we shall assume that
$\ellnp=0.4$ to illustrate the effects of interactions.

We consider in detail adding a cubic term to the loop potential,
$\kappa_3 \neq 0$.  We have also considered adding higher terms,
such as a quartic term, $\kappa_4 \neq 0$, but found numerically
that the results are rather similar.  
With  $\kappa_3 \approx 0.146$, 
we obtain $\ellnp \approx 0.4$ for the triplet loop.
The effective potential is shown in 
Fig.~\ref{fig:VeffN=3}, and 
exhibits an even smaller barrier between the two phases than
for $\kappa_3 = 0$.  

Numerically, we also find that at the transition, the
masses in the two minima are equal, to within a few percent,
for both $\kappa_3=0$ and $=0.146$.
This allows us to make the following observation.  Any polynomial
approximation to the potential fails for $\ell \approx 1$, due to
a logarithmic term which represents eigenvalue repulsion.
However, we can always use a polynomial approximation for small $\ell$.  
Consider a polynomial in $\ell$ to quartic order,
of the form $\sim \ell^2 (\ell - \ellnp)^2$: this represents
two degenerate minima, at $\ell = 0$ and $\ell = \ellnp$.
We find that at the transition, such a form
is approximately valid for $\ell \leq 0.6$.  Further, it is clear
that this potential gives equal masses in both the confined
and deconfined phases, which is nearly true numerically.

Such a quartic parametrization of the effective potential is, in fact, the
basis for the Polyakov loop model
\cite{banks_ukawa,loop1,loop2a,loop2b,loop2c,loop2d,loop2e,loop2f,loop3,sannino,ogilvie}.
The value
of $\ellnp$ in the Polyakov loop model, $\approx .55$ 
\cite{loop2a}, is close to that found from
the lattice~\cite{dhlop,bielefeld}.  Moreover, the form of 
the potential is graphically very similar to that found in the
matrix model, with an extremely small barrier between the two phases:
see Fig.~1 of~\cite{loop2a}.  

We also computed the effective mass in the matrix model.
At infinite $N$, the value about the Gross--Witten point is
given in eq.~(\ref{phys_mass_deconf}). 
For $N=3$, the effective mass is obtained numerically
by computing the curvature of $\veff$
about the non-trivial minimum $\ell_0$, cf.\ eq.~(\ref{define_eff_mass}). 
The results for $\kappa_3=0$ and $\kappa_3 \neq 0$ are shown
in Fig.~\ref{fig:mass} as a function of the expectation value of the
fundamental loop. At the Gross-Witten point, 
$\ell_0(T_d^+)= \frac{1}{2}$ and
$m^2_{eff}(T_d^+)=0$, as discussed in Sec. \ref{gross_witten_point}. 
When $N=3$, 
at the transition the effective mass is small, but nonzero.  In agreement
with the discussion of $\veff$ above, 
when $\kappa_3 \neq 0$, both the expectation value of the triplet
loop, and the effective mass, are smaller than for $\kappa_3 = 0$.
Except for this small effect, for both values of $\kappa_3$
the effective mass for $N=3$ is close to that at infinite $N$.  
This happens because the Vandermonde potential at $N=3$ is close
to $N=\infty$, and $\kappa_3 = 0.146$ is a small value.  For
example, since $\ell \approx \frac{1}{2}$ and $m^2 \approx 1$
at the transition, the cubic coupling,
$\kappa_3 \ell^3$,  is $\approx 8\%$
relative to the adjoint loop, $m^2 \ell^2$.

To compute the expectation values of loops in higher
representations we use a mean field analysis, 
like that of Sec.~\ref{sec_mean_field} \cite{dhlop}.
We checked that for
$\kappa_3 = 0$, the expectation value of the triplet loop, as
computed from the effective potential, agrees with the
mean field result.
The expectation values of the loops in the 
triplet, sextet, octet, and decuplet 
representations of $SU(3)$, computed in this way,
are shown in Fig.~\ref{fig:vevs_kappa0}.
We plot them as
functions of the coupling $m^2$, divided by the critical value
at which the transition occurs:
$m^2_{\rm crit} \approx 0.91$ for $\kappa_3 = 0$, and
$m^2_{\rm crit} \approx 0.97$ for $\kappa_3 = 0.146$.
For $\kappa_3 \neq 0$, the cubic term in the loop potential
reduces the expectation values of all loops, not just 
that of the triplet loop. 
When $m^2 \rightarrow \infty$, the effect of the cubic interaction
diminishes, as the expectation values of all loops
approach that for $\kappa_3 = 0$.

In agreement with the original results of \cite{dhlop}, 
we find that the triplet loop, as measured on the lattice,
agrees approximately with that for $\kappa_3 = 0.146$, assuming
a linear relation between $m^2$ and the temperature about $T_d$.
The lattice data is not sufficiently precise to make
a detailed comparison, however.  In particular, as the loop
approaches one, this is not an accurate way of determining
the relationship between $m^2$ and the temperature.  We return
to this point shortly.

For representations beyond the triplet, the difference loop,
introduced in \cite{dhlop}, is the remainder
between the expectation value of the loop,
and the result expected in the large $N$ limit, which is
a product of fundamental (and anti-fundamental) loops:
\beq
\delta \ell_\calr = \langle \ell_{\cal R}\rangle - \langle
\ell_3\rangle^{p_+} \langle\ell_3^*\rangle^{p_-} \; .
\eeq
The integers $(p_+,p_-)$ are $(2,0)$ for the sextet,
$(1,1)$ for the octet, and $(3,0)$ for the decuplet representation.
The difference loops are plotted in Fig.~\ref{fig:diffloops_kappa0}.
Here, too, we find that the cubic interaction only matters near
the transition:
$\delta \ell_8$, $\delta \ell_6$
and $\delta \ell_{10}$  are slightly smaller
in magnitude near $T_d$ when
$\kappa_3>0$, relative to their values for $\kappa_3 = 0$.

In all cases, we find that the difference loops are much
smaller than those measured on the lattice, Fig.~9 of~\cite{dhlop}.
To make a detailed comparison, it is necessary to know
the precise relation between the parameter $m^2$ of the matrix
model and the temperature.  Even without this relation, however,
the sextet difference loop is about four times 
larger on the lattice than in the matrix model.  For the octet
difference loop, the lattice data exhibits a sharp spike
close to the transition, about six times larger than the value in
the matrix model. 
There is, as of yet, no data for the decuplet loop from the lattice.
To describe these behaviors in a matrix model, it will certainly
be necessary to include fluctuations, due to kinetic terms in the
action.  It is not apparent whether the lattice data can be fit with couplings
and kinetic terms whose coefficients are independent of temperature.

We conclude by discussing the relationship of our results to the
Polyakov loop model 
\cite{banks_ukawa,loop1,loop2a,loop2b,loop2c,loop2d,loop2e,loop2f,loop3,sannino,ogilvie}.
In this model, the pressure is assumed to be a potential for
$\ell$ times $T^4$.  As discussed above, this form does seem to
work well near the transition.  
In order to fit the pressure away from $T_d$, however, it is necessary
to assume that the variation of $m^2$ is not simply linear in the
temperature.  As shown in \cite{loop2d}, by temperatures of
$\approx 3 T_d$, the pressure is nearly a constant times $T^4$;
this requires that $m^2$ is nearly constant with respect to
temperature.  As noted above, when the value of the loop is near
one, and given the uncertainties in the lattice data, we cannot exclude
such a change in the relationship between $m^2$ and the temperature.

For $N=3$, assuming that the real part of the Polyakov loop
condenses, one can compute masses for both
the real and imaginary parts of the Polyakov loop \cite{loop2c}.  
The effective masses discussed above refer only to that for
the real part of the Polyakov loop.  This is a (trivial) limitation
of the matrix model, and is easiest to understand in the perturbative
limit, when $m^2 \rightarrow \infty$, so $\ell \rightarrow 1$.
In this limit, $\boldl \approx \bf{1} + \sigma$, where
$\sigma$ is an element of the Lie algebra, and the action is
$\sim m^2 \; \tr \sigma^2$.  Integrating over $\sigma$, 
by (\ref{part_back_h}) the effective mass is related to
$1/m^2_{eff} \sim \int d\sigma
\; (\tr (\sigma^2))^2 \exp(- m^2 \tr \sigma^2) $,
and gives $m^2_{eff}\sim m^4$.
This is the same behavior as at large $N$, as can be read off from
(\ref{phys_mass_deconf}) when $\ell_0 \rightarrow 1$.  
If we do the same for the imaginary
part of the loop, which starts as ${\rm Imag}( \tr \boldl) \sim
\tr \sigma^3$, we would conclude that its effective mass squared
is $\sim m^6$.  This is wrong,
however: in the full theory,
the ratio between the effective mass for the imaginary,
and real, parts of the fundamental loop must be $3/2$,
as it is in the gauge theory in the perturbative limit.
The problem, as noted by 
Brezin {\it et al.} \cite{brezin},
is simply that because there are no kinetic terms,
one cannot generally compute two point functions in a matrix
model, but only one point functions.
It is really exceptional that we could obtain even the mass for
the real part of the fundamental loop.  Of course, this is
automatically remedied by adding kinetic terms to the matrix
model.

\section{Conclusions} 
\label{sec_Summary}

In this paper we developed a general approach to the deconfining
transition, based upon a matrix model of Polyakov loops.  The
effective action for loops starts with a potential term.
The Vandermonde determinant, which appears in the measure of the
matrix model, also contributes a potential term.

At infinite $N$, the vacua are the stationary points of an effective
potential, the sum of loop and Vandermonde potentials.  Because
of the Vandermonde potential, there is a transition with just a mass
term, at the Gross--Witten point.  This transition is of first order,
but without an interface tension between the two phases.  In the space
of all possible potentials, which include arbitrary interactions of
the Wilson line, the Gross--Witten point is exceptional.  Away from
it, the transitions appear ordinary, either of first order, with a
nonzero jump of $\ell$ and nonvanishing masses, or of second order,
with vanishing masses and no jump in the order parameter.  (There are
also third order transitions, with no jump in the order parameter and
nonzero masses.)  Only by tuning an infinite number of couplings to
vanish does one reach the Gross--Witten point, where the order
parameter jumps, and yet masses are zero.

We investigated the $N=3$ matrix model about the Gross--Witten point,
including interactions.  We found that the $N=3$ Vandermonde potential
is very close to $N=\infty$, so that for small interactions, the $N=3$
effective potential strongly resembles that of infinite $N$.  
We found that in the $N=3$ matrix model, 
corrections to the large $N$ limit are small, $\sim 1/N$.

What is so surprising about the lattice data is that the
$SU(3)$ deconfining transition is well described by a matrix
model near the Gross--Witten point \cite{dhlop}.
In the confined phase, on the lattice
the expectation value of the $Z(3)$ neutral
octet loop was too small to measure; in a matrix 
model, it vanishes identically.
At the transition, the renormalized triplet loop 
jumps to $\approx 0.4$ ~\cite{dhlop,bielefeld},
which is close to the Gross--Witten value of $\frac{1}{2}$.
Further, masses associated with the
triplet loop --- especially the string tension --- do decrease
significantly near the transition~\cite{three}.  

Of course, the proximity of the $SU(3)$ lattice data to the 
Gross--Witten point could be serendipitous, due 
to $N=3$ being close to the second order transition of $N=2$.  
For $N \geq 4$, the lattice does find a first order 
deconfining transition \cite{four_colors,teper,meyer}.  
The lattice data also shows that at a fixed value 
of $T/T_d$, mass ratios increase with $N$.
In the confined phase, one can compare the ratio of the
temperature dependent string tension to its value at 
zero temperature.  This can be
done for $N=2$ \cite{two_colors_B}, $3$
\cite{three}, $4$ and $6$ \cite{teper,meyer}. For example, 
comparing $N=4$ and $6$, Fig.~5 of \cite{meyer} shows
that at fixed $T/T_d$,
this ratio of string tensions increases as $N$ does.
In the deconfined phase, 
Fig.~7 of the most recent
work by Lucini, Teper, and Wenger \cite{teper}
shows that the ratio of the Debye mass at $T_d^+$,
to $T_d$, is significantly larger for $N=8$ than for $N=3$.
Thus, the lattice data strongly indicates that the $N=\infty$
theory is not exactly at the Gross--Witten point; if it
were, at finite $N$
one would expect that mass ratios would decrease, as $T\rightarrow T_d$,
in a manner which is nearly $N$-independent.

A recent, lengthy calculation by Aharony {\it et al.} shows
that on a small sphere,
at infinite $N$ the deconfining
transition is of first order \cite{aharony3}.
If $g$ is the gauge coupling, then
as $N \rightarrow \infty$, $g^2 N$ is a number of order one,
while $g^2 N \ll 1$ on a small sphere. 
In this limit, the coefficient of the quartic coupling
in the loop potential, (\ref{four_point_potential}), is
$\kappa_4 \sim (g^2 N)^2$ \cite{aharony};
\cite{aharony3} finds that 
$\kappa_4 \approx - (1/10) (g^2 N/(4\pi))^2$.
Although it is difficult to know how to normalize the coefficient,
its value suggests that on a small sphere, 
the theory is near the Gross--Witten point.  
{\it Perhaps} this remains true in an infinite volume.

With optimism, then, we assume that 
in the sense of Sec.~\ref{couplings_largeN}, the deconfining transition 
at $N=\infty$ is close to, but not at, the Gross--Witten
point.  We suggest that at finite $N$, fluctuations can drive
the theory much closer to this point: that it is, in the
space of all possible couplings, an infrared stable fixed point.   
Couplings only flow due to fluctuations, however, and in matrix
models these are manifestly $\sim 1/N^2$.  Thus this can only happen in a 
region of temperature which shrinks as $N$ increases,
for $|T-T_d|/T_d \approx 1/N^2$.  

A comparison of lattice simulations at different values of $N$
provides some evidence for a critical window which narrows 
with increasing $N$. 
Consider the decrease of the string tension near $T_d$.
Define a temperature $T_{1/2}$ as the point 
at which the string tension is half its value at zero temperature,
$\sigma(T_{1/2})\equiv 0.5\,\sigma(0)$.  
The lattice finds that 
the corresponding reduced temperature,
$\delta t_{1/2} \equiv(T_d - T_{1/2})/T_d$, 
decreases as $N$ increases:
$\delta t_{1/2} \approx 0.2$ at $N=2$ (Fig.~3 of \cite{two_colors_B}), 
$\approx 0.08$ when $N=3$ (Fig.~5 of \cite{three}),
and $\approx 0.05$ for $N=4$ (Fig.~5 of \cite{meyer}).  
These three points can be fit with 
\beq
\delta t_{1/2} \approx 0.8/N^2 \; .
\eeq
This really is a narrow window, with $\delta t_{1/2} \approx 0.02$ for
$N=6$.
%When $N=6$, this formula gives $\delta t_{1/2} \approx 0.02$;
%\cite{meyer} is not close enough to $T_d$ to test this.
%If correct, this really is a narrow window:
%at $N=8$, the string tension is only half its value at
%zero temperature within $\approx 1\%$ of $T_d$.

Unfortunately, a loop model cannot predict 
the ratio of the string tension at $T_d$, to zero temperature, because
we do not know what $m^2$ corresponds to zero temperature.
For three colors, this ratio is $\approx 0.12$ \cite{three}.
A loop model can predict the ratio of the string tension,
to the Debye mass, at $T_d$.
At infinite $N$, as $T\rightarrow T_d$
the string tension vanishes as $\sigma(T) \sim |T- T_d|^{1/2}$,
while the Debye mass vanishes more slowly,
$m_{Debye} \sim |T- T_d|^{1/4}$
(\ref{phys_mass_confined}) and 
(\ref{phys_mass_deconf_limit}) \cite{dhlop}.
That $\sigma(T)/m_{Debye}(T) \sim |T_d - T|^{1/4}$ 
as $T \rightarrow T_d$ is unlike 
ordinary second order transitions,
where the analogous ratio is constant in a mean field approximation.
For three colors, $\sigma(T_d^-)/m_{Debye}(T_d^+) \approx 0.93\pm 0.02$
at the $N=3$ analogy of the Gross--Witten point, and $\approx 0.99 \pm 
0.03$ when
$\kappa_3 = 0.146$, Sec.~\ref{three_colors}.  For four colors,
it is $\approx 0.63 \pm 0.02$ 
at the $N=4$ analogy of the Gross--Witten point \cite{mean},
and decreases with increasing $N$.

The most direct way 
to test if the transition for $N \geq 4$ is close to the Gross--Witten
point is to compute the value
of the renormalized fundamental 
loop (very) near $T_d^+$ \cite{dhlop,bielefeld}, and see if it is
close to $\frac{1}{2}$.  The change in the masses
near $T_d$ is probably more
dramatic, though: since the latent heat grows as $\sim N^2$
\cite{teper,meyer}, such a decrease,
especially in a window which narrows as $N$ increases, would be 
most unexpected.

To rigorously 
demonstrate the connection between loop models and deconfinement,
it will be necessary to include
fluctuations at finite $N$.  At the beginning of Sec.
\ref{sec_InfiniteN}, we blithely ignored the kinetic terms
of Sec. \ref{matrix_models_loops}, proceeding immediately to a
mean field theory, which is an expansion in bare quantities.
Including fluctuations will require the analysis of the renormalized
theory.  Fluctuations are most important in the lower
critical dimension, which is two \cite{zinn,beta},
and is relevant for the deconfining transition in three spacetime
dimensions \cite{threedim}.
The effect of
fluctuations
can also be computed by working down from the upper critical
dimension, which is four.  In this vein, we comment
that the entire discussion of Sec. \ref{sec_InfiniteN}
is dominated by the Vandermonde potential, which arises from
the measure of the matrix integral.  
In the continuum limit, terms in the
measure can be eliminated from perturbation theory by the
appropriate regulator \cite{zinn}.
The measure can contribute non-perturbatively, though:
see, {\it e.g.},
Eq. (73) of Billo, Caselle, D'Adda, and Panzeri \cite{largeN}.
In general, it will be interesting to develop the renormalization group
for what may be a new universality class, about the Gross--Witten point.

\vspace{.25in}
{\bf Acknowledgements}
The research of A.D.\ is supported by
the BMBF and the GSI; 
R.D.P., by the U.S. Department of Energy grant DE-AC02-98CH10886;
J.T.L., by D.O.E.\ grant DE-FG02-97ER41027.
R.D.P. also thanks: the Niels Bohr Institute, for their gracious
hospitality during the academic year 2004-2005; 
the Alexander von Humboldt Foundation, for their support;
K. Petrov, for discussions about three dimensions;
P. Damgaard, for many discussions; 
and most especially, K.
Splittorff, for frequent discussions, and for collaborating on
part of the results in Secs. III and IV.

\vspace{.25in}
\newpage
\begin{figure}
\begin{center}
\epsfxsize=.48\textwidth
\epsfbox{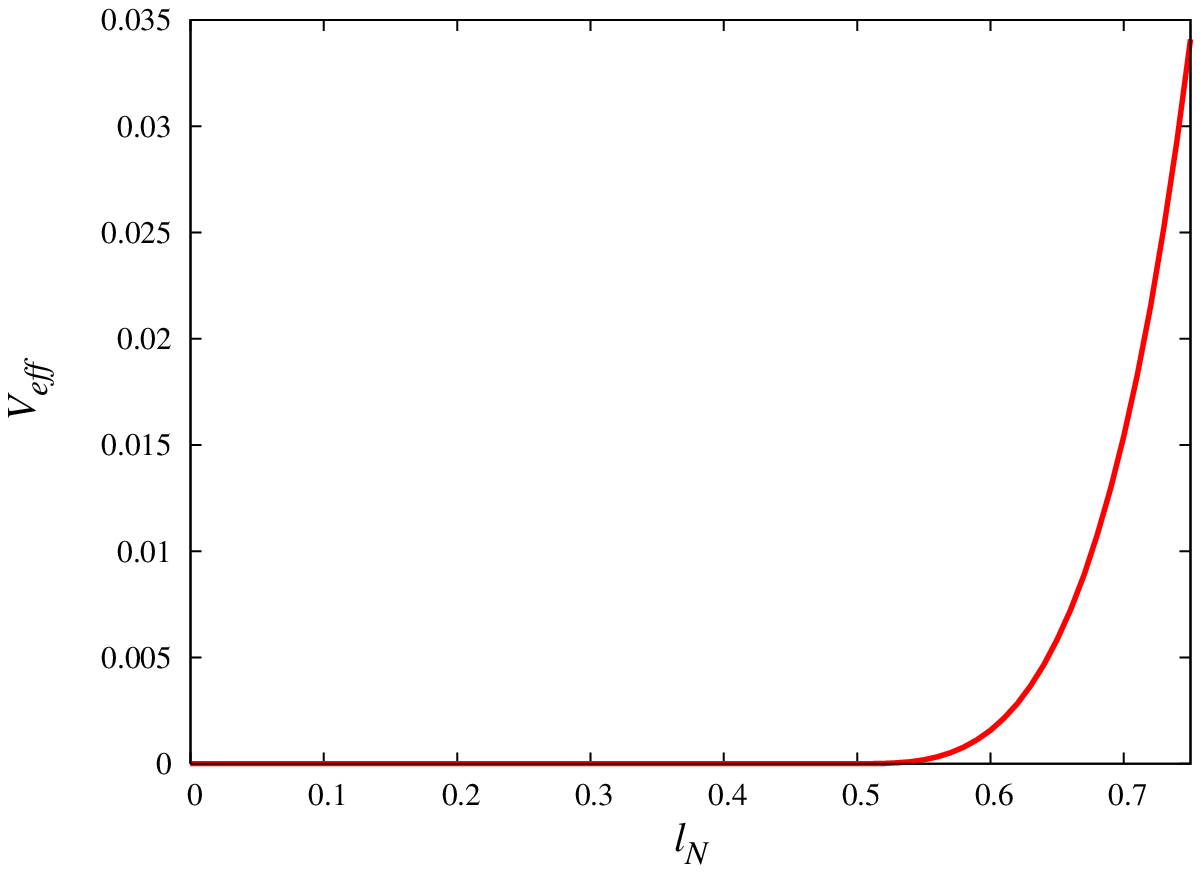}
\end{center}
\caption{The $N=\infty$ effective potential at the Gross--Witten point.
When $\ell_N < \frac{1}{2}$, the potential vanishes identically.
In Figs. 1, 2, and 3, there is a discontinuity, of third order,
at $\ell_N = \frac{1}{2}$, although it is only apparent here.}
\label{fig:gw}
\end{figure}

\begin{figure}
\begin{center}
\epsfxsize=.48\textwidth
\epsfbox{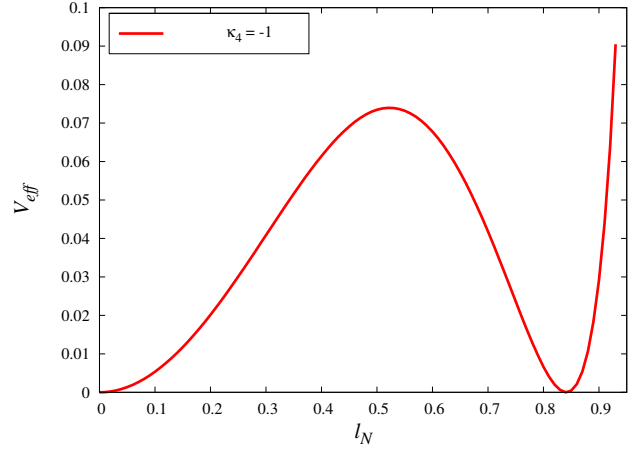}
\end{center}
\caption{The $N=\infty$ 
effective potential for $\kappa_4 = -1$, where the
transition is strongly first order.}
\label{fig:first}
\end{figure}

\begin{figure}
\begin{center}
\epsfxsize=.48\textwidth
\epsfbox{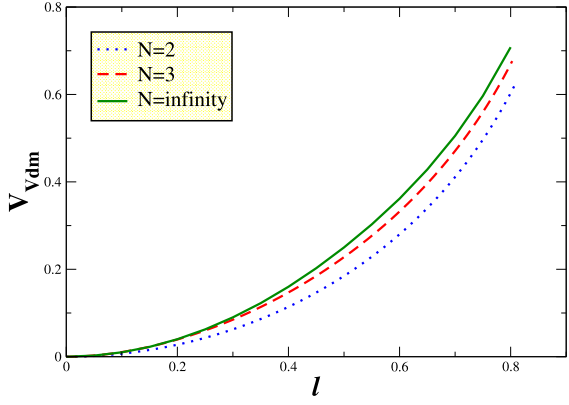}
\end{center}
\caption{The Vandermonde potential, ${\cal V}_{Vdm}$, for $N=2$
  (dotted line), $N=3$ (dashed line) and $N=\infty$ (full line).}
\label{fig:Vdm}
\end{figure}

\begin{figure}
\begin{center}
\epsfxsize=.48\textwidth
\epsfbox{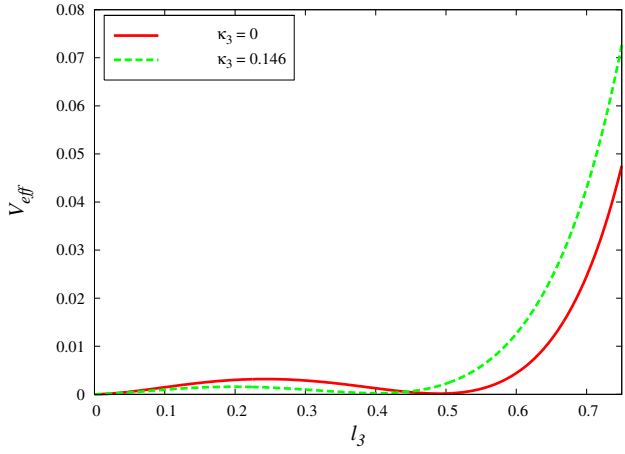}
\end{center}
\caption{The $N=3$ effective potential, at the transition,
for $\kappa_3=0$ (full line) and $0.146$ (dashed line).}
\label{fig:VeffN=3}
\end{figure}

\begin{figure}
\begin{center}
\epsfxsize=.48\textwidth
\epsfbox{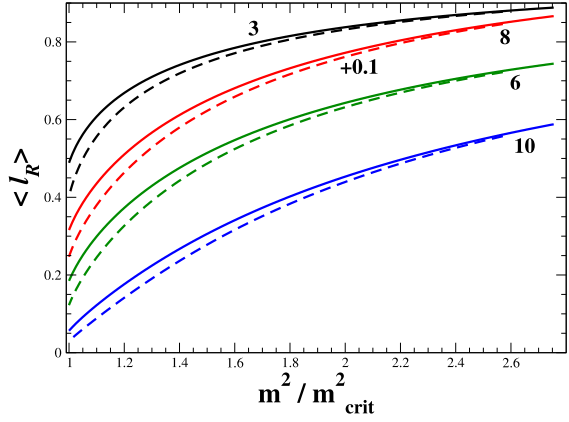}
\end{center}
\caption{Expectation values for the 
$3$, $6$, $8$, and $10$ representations of $SU(3)$, for
$\kappa_3 = 0$ (full lines) and $0.146$ (dashed lines),
as a function of the ratio of $m^2$ to its critical value.  
For better visibility, the expectation value of the octet has
been shifted by +0.1}
\label{fig:vevs_kappa0}
\end{figure}

\begin{figure}
\begin{center}
\epsfxsize=.48\textwidth
\epsfbox{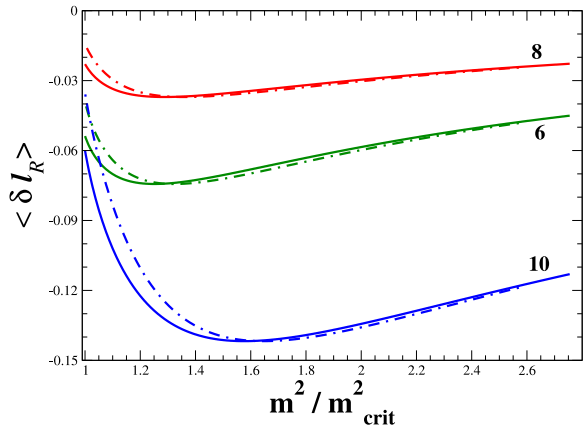}
\end{center}
\caption{Difference loops for $6$, $8$, and $10$ representations
of $SU(3)$, for $\kappa_3 = 0$ (full lines) and $0.146$ (dashed lines),
as a function of the ratio of $m^2$ to its critical value.}
\label{fig:diffloops_kappa0}
\end{figure}

\begin{figure}
\begin{center}
\epsfxsize=.48\textwidth
\epsfbox{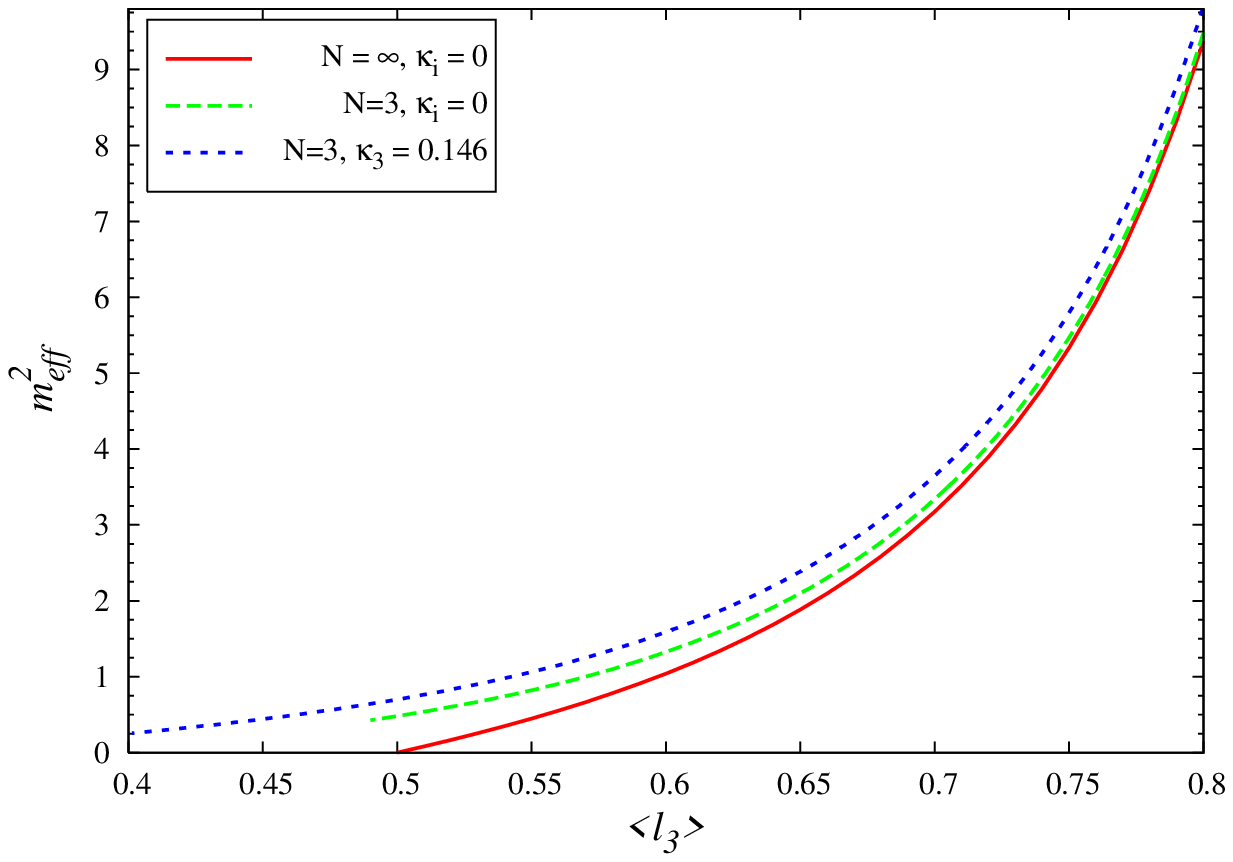}
\end{center}
\caption{The effective mass squared, for the fundamental loop in the
deconfined phase,
versus its expectation value.  The three curves
are $N=\infty$ about the Gross-Witten
point, all $\kappa_i = 0$ (full line);
$N=3$ with all $\kappa_i=0$ (dashed line);
and $N=3$ with $\kappa_3=0.146$ (dotted line).}
\label{fig:mass}
\end{figure}


\begin{thebibliography}{999}
\bibitem{difference}
For spatial separation
$R$, the renormalization constant for the Wilson loop at $T\neq 0$ is
a product of two terms, one for mass renormalization of the loop,
$\calz_{line}$, and one involving the cusps in the loop, $\calz_{cusp}$.
The first involves 
the product of the divergent mass times the total length of
the loop: $\calz_{line} = \exp(- 2 m^{div}(R + 1/T))$,
where $m^{div} \sim 1/a$ at lattice spacing $a$ \cite{dhlop}.
The latter involves the anomalous dimensions for the
four cusps in a rectangular loop \cite{cusp}:
$\calz_{cusp} = \exp( - 4 \int dg \; \gamma(\pi/2,g)/\beta(g) )$,
where $\gamma(\pi/2,g)$ is the anomalous dimension for a cusp
of $90^o$, 
$\beta(g)$ is the $\beta$-function, and $g$ is the gauge
coupling constant.  There is no condition needed to fix $\calz_{line}$
at some scale, but one is needed for $\calz_{cusp}$ \cite{dhlop,cusp}.
Straight Polyakov loops don't have cusps 
(although one can construct ``crooked'' Polyakov loops
which do \cite{dhlop}).  The contribution of the anomalous dimensions
for the four cusps may explain the persistent discrepancy, 
of $\sim 10\%$, between the methods of \cite{dhlop},
for the renormalized Polyakov loop, and that of
\cite{bielefeld}, for the renormalized Wilson loop at $T \neq 0$.
\end{thebibliography}
\end{document}